\documentclass[aps,prstab,superscriptaddress,twocolumn,english,longbibliography]{revtex4-1}
\usepackage[utf8]{inputenc}
\usepackage{lineno}
\usepackage{grffile}
\usepackage{graphicx}
\graphicspath{{./figs/}}
\usepackage{subcaption}

\usepackage[bookmarks, colorlinks=true, linktoc=page, pdftex, linkcolor=black, citecolor=black, urlcolor=black]{hyperref}
\usepackage[all]{hypcap} 
\usepackage{siunitx}
\usepackage{multirow}
\usepackage{placeins}
\usepackage{xspace}
\usepackage{verbatim}
\usepackage{upgreek,textgreek}
\usepackage{url}
\usepackage{engord}

\usepackage{newunicodechar}
\newunicodechar{α}{\ensuremath{\alpha}}
\newunicodechar{β}{\ensuremath{\beta}}

\newcommand{\betastar}{\ensuremath{\beta^*}\xspace}
\begin{document}
\title{Performance of the Large Hadron Collider cleaning system during the squeeze: simulations and measurements}

\author{S. Tygier}
\affiliation{The University of Manchester, Oxford Road, Manchester M13 9PL, UK}
\email{sam.tygier@manchester.ac.uk}
\affiliation{The Cockcroft Institute, Daresbury, WA4 4AD, UK}
\author{R.B. Appleby}
\affiliation{The University of Manchester, Oxford Road, Manchester M13 9PL, UK}
\affiliation{The Cockcroft Institute, Daresbury, WA4 4AD, UK}
\author{R. Bruce}
\affiliation{CERN, Geneva, Switzerland}
\author{D. Mirarchi}
\affiliation{CERN, Geneva, Switzerland}
\author{S. Redaelli}
\affiliation{CERN, Geneva, Switzerland}
\author{A. Valloni}
\affiliation{The University of Manchester, Oxford Road, Manchester M13 9PL, UK}
\affiliation{CERN, Geneva, Switzerland}

\begin{abstract}
The Large Hadron Collider (LHC) at CERN is a 7~TeV proton synchrotron, with a design stored energy of 362~MJ per beam. The high-luminosity (HL-LHC) upgrade will increase this to 675~MJ per beam. In order to protect the superconducting magnets and other sensitive equipment from quenches and damage due to beam loss, a multi-level collimation system is needed. Detailed simulations are required to understand where particles scattered by the collimators are lost around the ring in a range of machine configurations. Merlin++ is a simulation framework that has been extended to include detailed scattering physics, in order to predict local particle loss rates around the LHC ring. We compare Merlin++ simulations of losses during the squeeze (the dynamic reduction of the $\beta$-function at the interaction points before the beams are put into collision) with loss maps recorded during beam squeezes for Run 1 and 2 configurations. The squeeze is particularly important as both collimator positions and quadrupole magnet currents are changed. We can then predict, using Merlin++, the expected losses for the HL-LHC to ensure adequate protection of the machine.
\end{abstract}

\maketitle

\section{Introduction}
\label{sec:intro}

\begin{figure}[!htbp]
	\begin{center}
		\includegraphics[width=1.0\columnwidth]{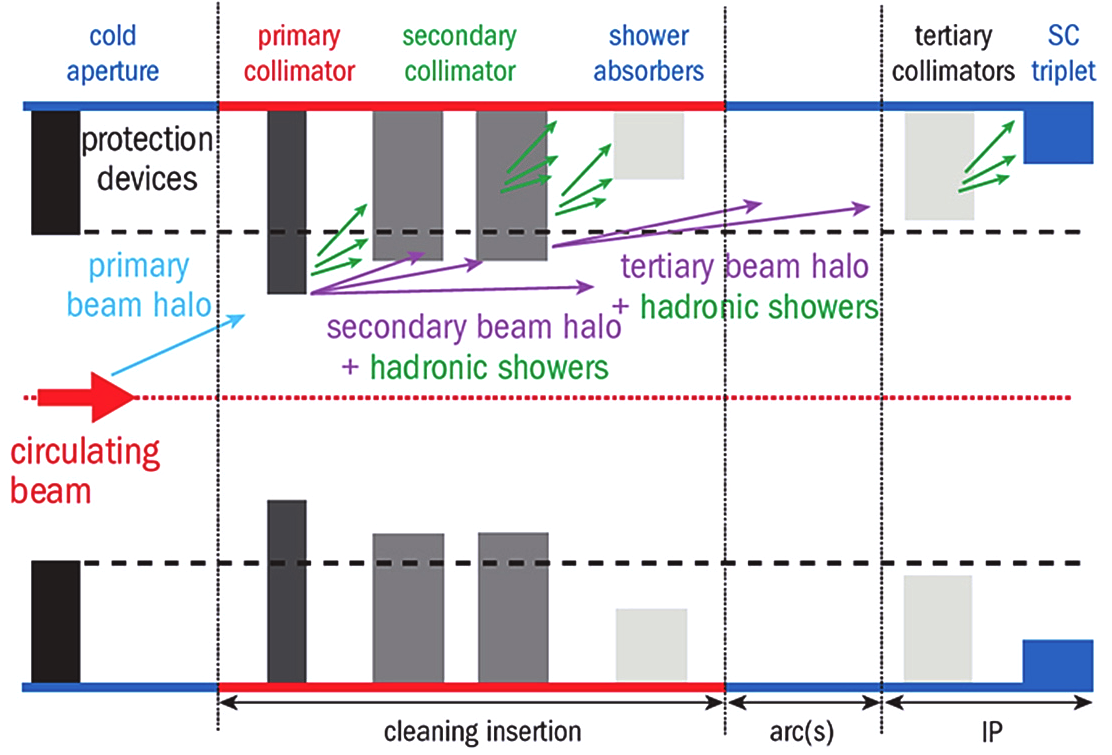}
		\caption{Overview of the main components involved in the multistage collimation system.}
		\label{fig:col_sys_hllhc}
	\end{center}
\end{figure}

The LHC collimation system~\cite{bruning_lhc_2004, assmann_final_2006} is designed to protect the ring from normal beam losses caused by diffusion and scattering, as well as abnormal fast losses. In order to achieve this it uses a multi-stage system for betatron cleaning, installed in Insertion Region 7 (IR7), and a similar but reduced system for momentum cleaning installed in IR3. The main components are shown in Fig.~\ref{fig:col_sys_hllhc}. The primary collimators (TCP) are made of carbon-fiber composites (CFC), and sit closest in to intercept the beam halo. The secondary collimators (TCS) also CFC and shower absorbers (TCLA) made of tungsten absorb the deflected protons and secondary particles.
There are also tungsten tertiary collimators (TCT) that provide extra protection for the experiments, as well as TCLs to absorb collision debris.
About 150 m upstream of each experiment, a pair of one horizontal (TCTPH) and vertical (TCTPV) TCT is installed. 
The LHC's multi-stage collimation system has proved to be performing reliably during Run 1 (2010-2012) with beam energies of 3.5~TeV and 4~TeV and the start of Run 2 (2015-2018) at 6.5~TeV. However, the future physics program at 7~TeV and the higher intensities of HL-LHC~\cite{apollinari_high-luminosity_2017, bruning_high_2015} provide new challenges, so it is important that the performance is well understood and can be accurately simulated. The \mbox{HL-LHC} program includes upgrades to the collimation system such as new collimators, new jaw materials~\cite{borg_thermostructual_2018} and new embedded instrumentation~\cite{valentino_final_2017}.

In this article we present predictions for HL-LHC losses during luminosity levelling, as well as measurements of the performance of the LHC collimation system during Run 1 and 2.
Previously, studies of LHC collimation efficiency have been successfully performed using the SixTrack code with comparisons to BLM data at fixed optical configurations~\cite{schmidt_sixtrack_1994,noauthor_sixtrack_2018,bruce_sources_2013,robert-demolaize_new_2005,bruce_simulations_2014,bruce_cleaning_2014,hermes_measured_2016,mirarchi_cleaning_2016,bruce_reaching_2017}. These demonstrate that maps of proton loss locations are useful for evaluating collimation scenarios and as input to energy deposition studies using codes such as FLUKA~\cite{bohlen_fluka_2014}.  In this article we use instead the code Merlin++, which is described in section~\ref{sec:sims} and present the first comparison loss maps taken during the squeeze.
In section~\ref{sec:loss} the Beam Loss Monitors (BLMs)~\cite{holzer_beam_2005, holzer_development_2008} are described. These are used to validate the simulation code. This system consists of about 4000 ionisation chambers distributed around the ring to monitor losses at critical elements.

For this work we consider the slow losses that occur during normal operation of the LHC. Particles in the core of the beam can be excited to higher amplitudes by a number of effects, drifting out to form the beam halo. When a particle's amplitude is large enough, it is intercepted by the collimators.

\begin{figure}[!htbp]
	\begin{center}
		\includegraphics[width=1.0\columnwidth]{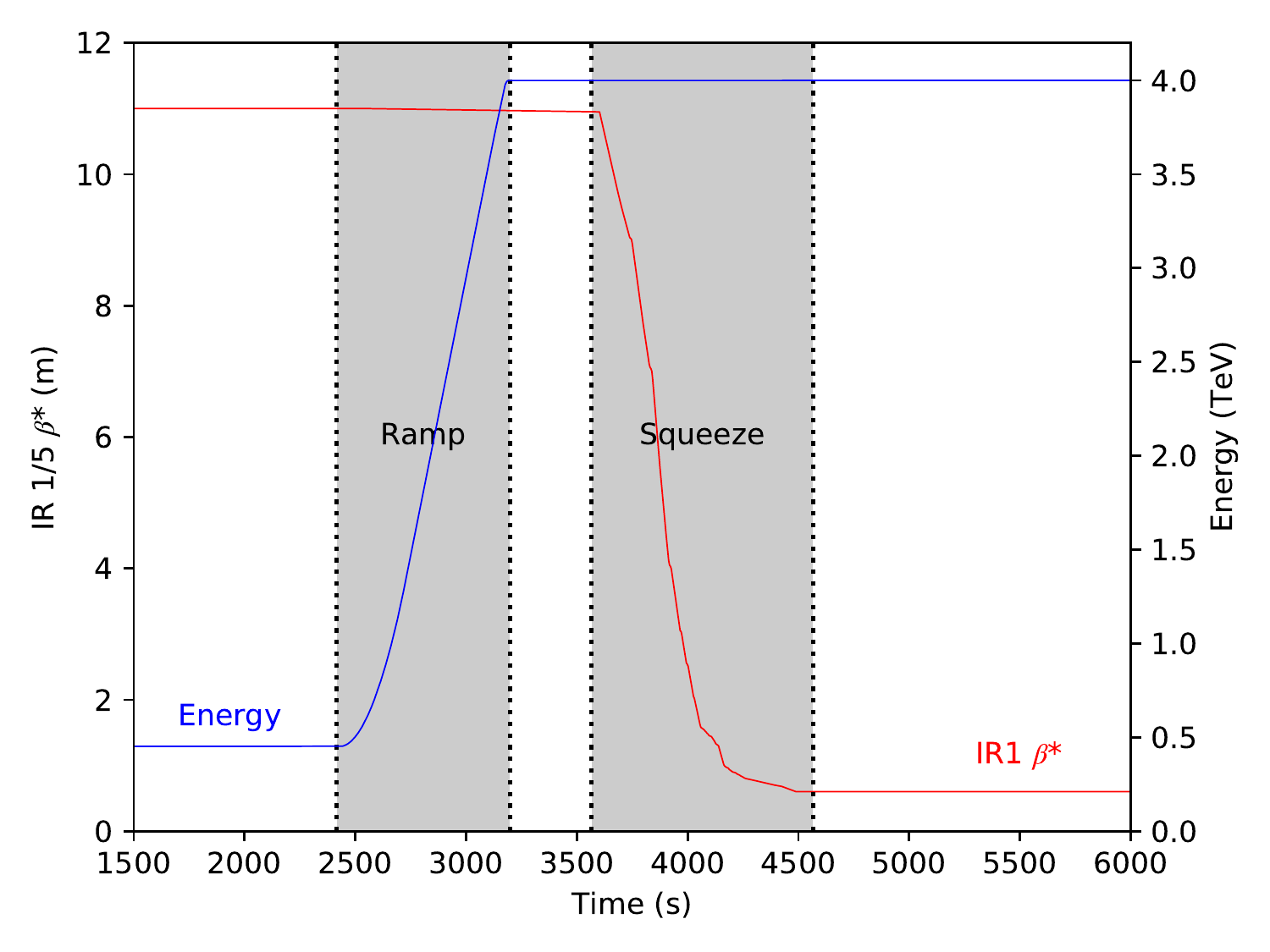}
		\caption{\betastar in IR 1/5 and beam energy during a typical squeeze during 2012, showing the ramp and squeeze periods.}
		\label{fig:beam_modes}
	\end{center}
\end{figure}

Before bringing the LHC beams into collision, they must first be ramped from injection energy (450 GeV) to full energy (6.5 TeV in Run 2) and the \betastar (the $\beta$-function at the experiment interaction points (IPs)) reduced. This latter part of the operational cycle is called the squeeze. In Run 1 these actions were performed separately, however during Run 2 a combined ramp and squeeze sequence was introduced to reduce the cycle duration. Figure \ref{fig:beam_modes} shows the beam modes for a typical Run 1 production fill, from injection of the physics beam through to stable beams for physics production. In the Fig.~\ref{fig:beam_modes} the squeeze begins at around 3500 seconds, with the \betastar at IP~1/5 being reduced from 11~m to 0.6~m.

The squeeze is an important time for the collimation system as there are dynamic changes to the machine optical configuration and collimator positions while the stored energy in the beam is at its maximum. It also provides a good opportunity to validate simulation against measurements in a range of configurations, allowing investigation into any differences found.

Sections \ref{sec:run1} and \ref{sec:run2} compare the simulations to data for Run 1 and 2 respectively. This gives us the confidence in Merlin++'s particle tracking and scattering models, in order to use it for making predictions of future configurations.
In section~\ref{sec:hl-lhc} we evaluate the loads on the collimators in the HL-LHC, for the most pessimistic loss scenario allowed at full energy. This corresponds to a 0.2~h beam lifetime over 10~s~\cite{redaelli_cleaning_2015,assmann_preliminary_2002}, giving a total loss power of about 1 MW.

\section{Calculation of beam losses}
\label{sec:sims}

To calculate the losses in an accelerator we must model the trajectories of particles in the magnetic lattice and also the passage and scattering of particles in the materials that make up the collimators.

Merlin++~\cite{merlin_developers_merlin++_2018}, previously known as MERLIN, is a modular object-oriented accelerator simulation framework, featuring 6D thick lens tracking. Initially developed for the International Linear Collider's beam delivery systems~\cite{krucker_ilc_2006}, it has since been extended to support synchrotrons. It is written in C++ and can be easily extended by the user, for example to add new physics models. It has multiprocessor support using the MPI protocol for communication; however for non-collective effects as used in collimation studies, it is more convenient to run multiple independent processes and sum the results.

The user-created program calls the Merlin++ library in order to define a beam line, add appropriate physics processes, create a beam and then initiate tracking. In these studies the MADInterface module was used to read in a lattice description from a MADX~\cite{cern_madx_2018} optics calculation. Machine apertures and collimator gaps were similarly defined in separate files.

While the codes SixTrack~\cite{schmidt_sixtrack_1994} and FLUKA have been used for similar comparisons to BLM data previously~\cite{bruce_simulations_2014} it is useful to have multiple independent simulation codes in order increase confidence and constrain systematic uncertainness. Merlin++ features thick lens tracking compared to SixTrack's thin lens method and more advanced scattering model as described below. Merlin++ also has some technical advantages, for example its modular C++ design simplifies the incorporation of new physics models

The scattering physics used to model the passage of protons through material has recently been upgraded~\cite{molson_proton_2014}. It contains advanced models of the following processes: multiple Coulomb scattering; ionisation based on Landau theory; Rutherford scattering; a new elastic scattering model; and a new single-diffraction dissociation model. The model uses proton-nucleon scattering based on the Donnachie and Landshoff (D-L) description of Pomeron and Reggeon exchange and has been fitted to elastic and diffractive scattering data from a large number of previous experiments. The development and implementation of these models are described in detail in our earlier paper~\cite{appleby_practical_2016}. This model has now been included into the Pythia event generator, under the name ABMST model~\cite{rasmussen_models_2018}. These new scattering models give different predictions for losses around the ring compared to the K2 model in older versions of SixTrack, specifically the single diffractive model was found to cause a significant shift of losses from cold to warm regions. For performance reasons, only the leading proton from each interaction is modelled; equivalent to assuming that the secondaries deposit their energy close to the interaction.

To simulate loss maps we use an approach similar to earlier LHC studies \cite{robert-demolaize_new_2005, bruce_simulations_2014}. We track $10^8$ protons for 200 turns. The initial particle bunch is generated at the face of the primary collimator in the excitation plane. This simulated bunch is a ring in phase-space in the excitation plane that intersects the collimator jaw by 1 \si{\micro m} from its edge, the impact parameter, and Gaussian in the opposite plane. It is pre-filtered so that every proton interacts with the primary collimator on the first turn. This saves significant computer resources compared to modelling the real LHC bunch distribution and the diffusion of particles from the core of the bunch. As the beam is tracked, it is compared to the machine aperture at each element. If a proton hits the aperture in a collimator then it is scattered according to the process cross sections. If the proton hits the aperture of other lattice elements, it is considered as lost at this location. The loss map records the location of every proton loss with resolution of 10~cm.

The primary collimators are set to be the tightest aperture restriction in the machine, so that they are the first material that a proton with sufficient amplitude will hit. The protons that scatter without being absorbed will go on to hit other elements in the ring. If all protons hitting the primary collimator were absorbed by the collimation system we would consider it to have 100~\% cleaning efficiency. In practice we typically see about 8~\% of protons absorbed during a single pass through a TCP due to inelastic collisions or losing over 95~\% of their initial energy which is considered to be a loss. Over subsequent turns around the ring most protons will be lost in the TCPs. To produce a loss map, which shows the distribution of losses around the ring, we measure the proton loss locations and use them to calculate the local cleaning inefficiency. The simulated local cleaning inefficiency, $\eta_{\mathrm{loc}}$, is given by the ratio of particles lost on a given section, $N_{\mathrm{loc}}$, (either an element or bin along the S coordinate) to particles lost in the primary collimators, $N_{\mathrm{tot}}$, normalised to the length of the section, $\Delta s$, to make the value independent of bin size i.e.
\begin{equation} \label{eq:scaling_law}
\eta_{loc} = \frac{N_{loc}}{N_{tot} \Delta s}.
\end{equation}
The local cleaning inefficiency can then be multiplied by the total beam loss rate to find the local proton loss rate.

\section{Loss measurements}
\label{sec:loss}

The LHC BLM system uses ionisation chamber charged particle detectors to measure the radiation levels around the LHC ring~\cite{holzer_beam_2005, holzer_development_2008}. They are used during operation to trigger a beam dump if loss thresholds are exceeded. They also provide continuous measurements of normal beam loss around the ring during the LHC operations and are used to record beam loss during the validation campaigns of the collimation system, when artificial losses are induced with safe low intensity beams to assess the system response.

To generate a loss map one of the beams is excited in a given plane using the transverse dampers (ADTs) and the losses are recorded~\cite{hofle_controlled_2012}. This allows a clean loss map for an individual beam and plane to be made. Measured LHC loss maps have been studied previously in~\cite{bruce_simulations_2014, bruce_reaching_2017}.

During 2012 several loss maps were recorded at 4~TeV in the flat top and fully squeezed optics configurations. No deliberate loss maps were made with the intermediate squeeze optics at 4~TeV, but as the BLM signals are recorded continuously it is possible to look at the natural losses during the squeeze. In 2016 at 6.5~TeV, loss maps were generated at the intermediate squeeze points as well as the end points.

These loss maps are crucial to validate simulations, to ensure a good understanding of the collimation system, and hence demonstrate the performance of the HL-LHC layout. In the following sections we show the validation for Run~I and 2 of the LHC.

\section{Run 1 - 4 TeV}
\label{sec:run1}

\subsection{Run 1 LHC 2012 configuration}

\begin{figure}[!htbp]
	\begin{center}
		\includegraphics[width=1.0\columnwidth]{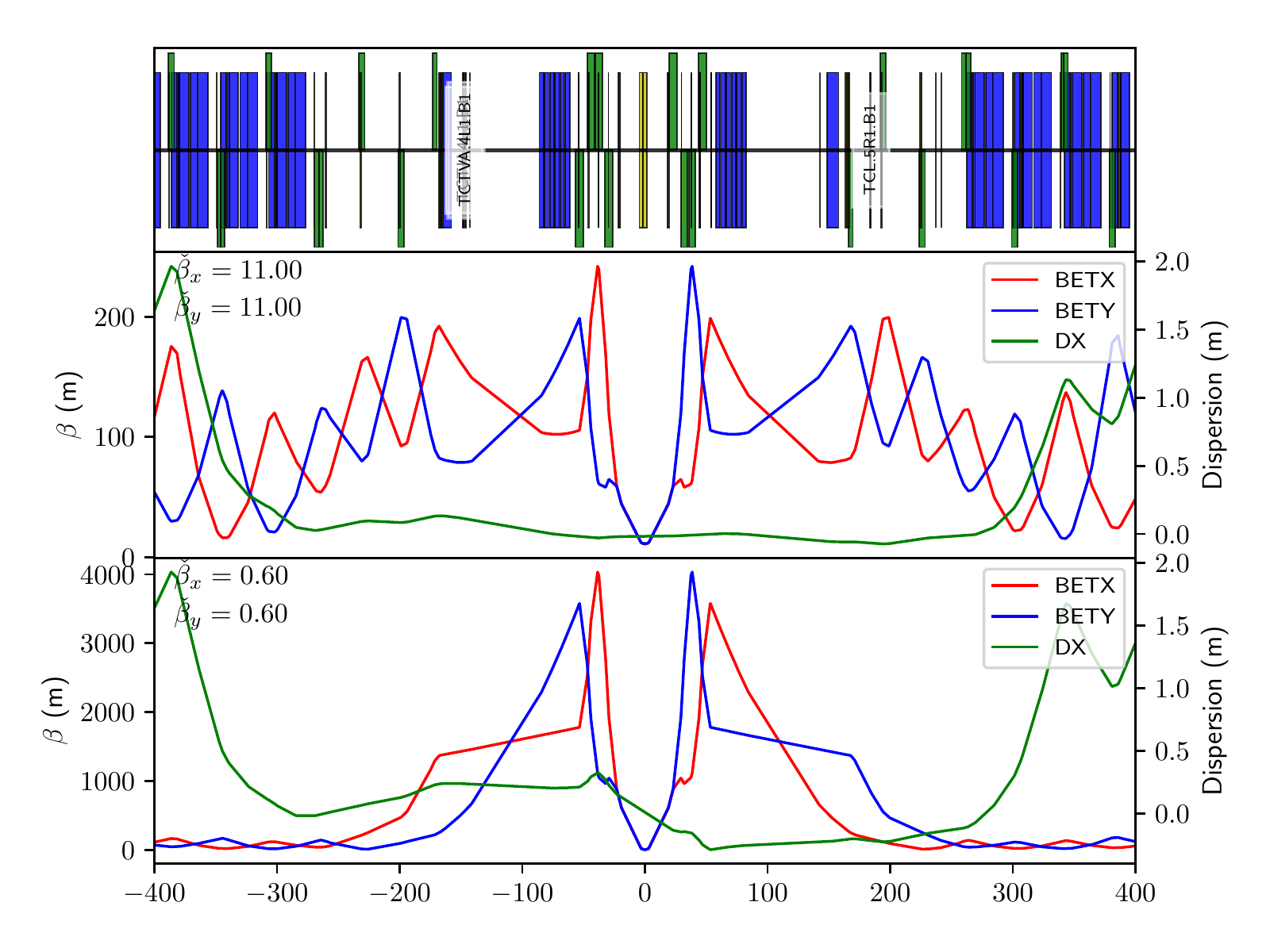}
		\caption{$\beta$-functions and dispersion at the ATLAS experiment for flat top (top) and squeezed (bottom) settings for the 2012 run configurations.}
		\label{fig:atlas_sqeeze}
	\end{center}
\end{figure}

In this article we investigate losses at intermediate optics points while the machine is in squeeze mode. Here the optics configurations are changing as a function of time, see Fig.~\ref{fig:beam_modes}. As \betastar is reduced, the $\beta$-function in the inner triplets must increase, as shown in Fig.~\ref{fig:atlas_sqeeze}. Table \ref{tab:optics2012} shows the optics settings for the squeeze during 2012.

\begin{table}[!htbp]
\caption{Optics settings for squeeze in 2012 at 4~TeV. Crossing angle can be in the horizontal (H) or vertical (V) plane. Note that for IR2 and 8, the external crossing angle, applied on top of the spectrometers and their compensation bumps, is given.}
\centering
\begin{tabular}{|l|l|l|}
\hline
  & \betastar &   half crossing angle \\
  &        (m)& (\si{\micro rad})\\
\hline
ATLAS (IP1) & 11 $\rightarrow$ 0.6 & -145 V \\
CMS   (IP5) & 11 $\rightarrow$ 0.6 & 145 H \\
ALICE (IP2) & 10 $\rightarrow$ 3.0 & -90 V \\
LHCb  (IP8)& 10 $\rightarrow$ 3.0 & -220 H \\
\hline
\end{tabular}
\label{tab:optics2012}
\end{table}

As well as the optics configuration changing during the squeeze, the TCTs are also in motion. The collimators are set in units of beam sigma, so as the beam envelope at the collimator's location changes, the collimators jaws are adjusted. Also the TCTs at the experiments are only brought into their tightest position during the squeeze as the normalised triplet aperture reduces, so their jaw gap in sigma units is decreased during the squeeze, as shown in Table~\ref{tab:collimators2012}. We therefore concentrate this study on the losses at the TCTs.

\begin{table}[!htbp]
\caption{Collimator settings for squeeze in 2012 at 4~TeV, from 11~m to 0.6~m at IP 1/5. Using a normalised beam emittance of 3.5~\si{\micro m}}
\centering
\begin{tabular}{|l|l|l|}
\hline
Region & Type & Gap ($\upsigma$)\\
\hline
IR7 & TCP & 4.3 \\
IR7 & TCS & 6.3 \\
IR7 & TCLA & 8.3 \\
\hline
IR3 & TCP & 12.0 \\
IR3 & TCS & 15.6 \\
IR3 & TCLA & 17.6 \\
\hline
IR1 & TCT & 26.0 $\rightarrow$ 9.0\\
IR5 & TCT & 26.0 $\rightarrow$ 9.0\\
IR2 & TCT & 26.0 $\rightarrow$ 12.0\\
IR8 & TCT & 26.0 $\rightarrow$ 12.0\\
\hline
\end{tabular}
\label{tab:collimators2012}
\end{table}

\subsection{Simulated loss maps}

Loss maps were simulated in Merlin++ at 8 points within the squeeze covering \betastar at IP1 and IP5 from 11~m to 0.6~m. A bunch of $10^8$ protons were tracked for 200 turns. This is sufficient to give good statistics for all relevant collimator and ring losses, even for particles that survive for multiple turns after their first scatter.

\begin{figure*}[!htb]
	\begin{center}
		\includegraphics[width=0.85\textwidth]{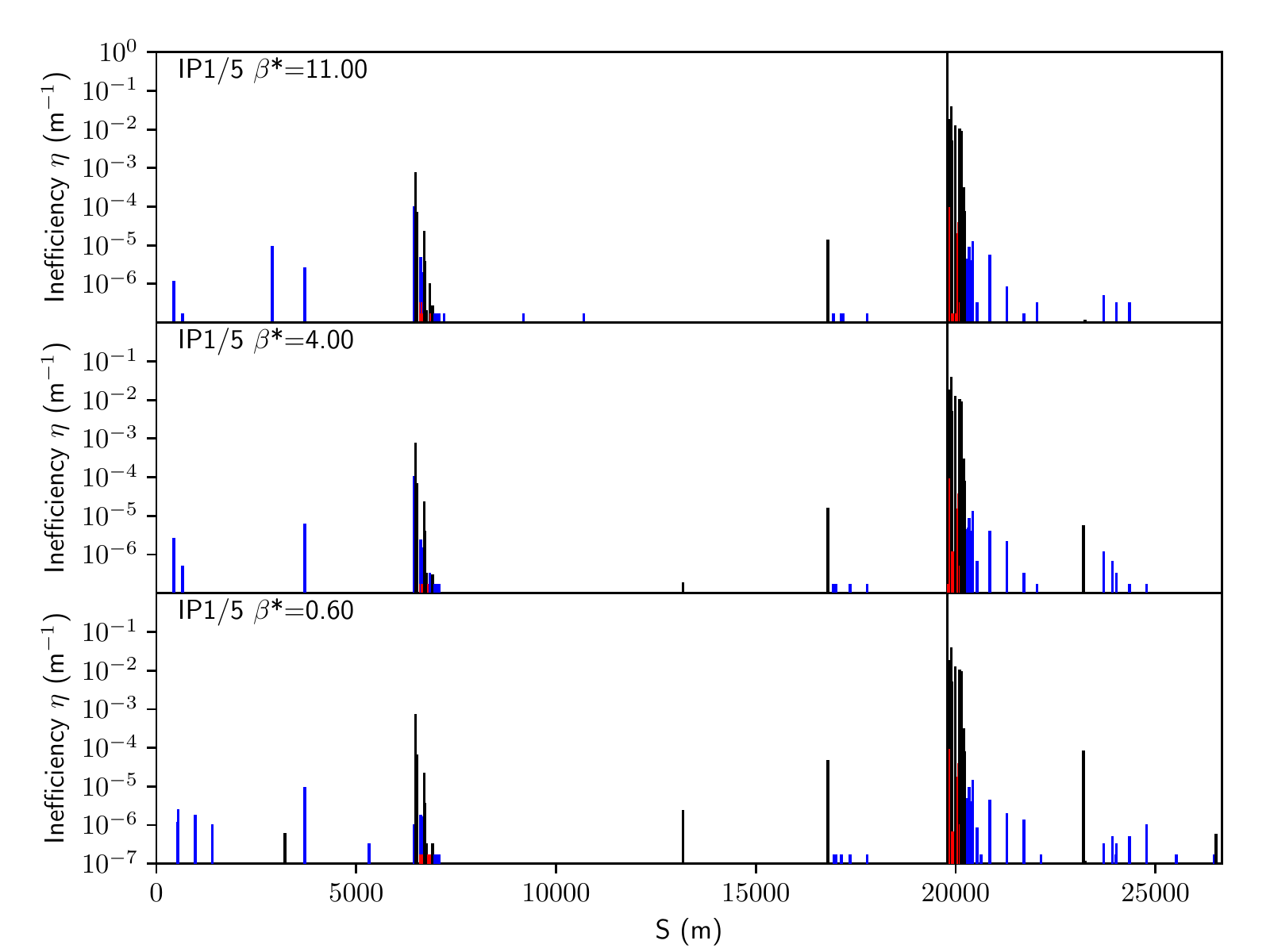}
		\caption{Loss maps, as simulated with Merlin++ for initial losses in the horizontal plane, for Beam 1. Made at different points in the squeeze with \betastar of 11, 4, and 0.6~m, for the 2012 4~TeV configuration. Losses in collimators are shown in black, in cold magnets in blue and in warm magnets in red.}
		\label{fig:4tev_merlin_lossmap}
	\end{center}
\end{figure*}

Figure \ref{fig:4tev_merlin_lossmap} shows examples of the loss maps at 3 of the optical configurations.
The highest losses occur on the TCPs in IR7 at around 20000~m from IR1 as expected, then lower losses along the cleaning hierarchy. Also significant losses at the momentum cleaning collimators in IR3, at around 7000~m, are observed.
The losses at the TCTs in front of the experiments in IR1/2/5/8 do not appear until the later stages of the squeeze.
The main cold losses are in the IR7 dispersion suppressor, which is the bottleneck in terms of local cleaning inefficiency.

\subsection{Measured squeeze losses in 2012}

As there were no dedicated loss maps made during 2012 at the intermediate squeeze optics settings, we compare to the BLM measurements made during normal LHC operation. These have a number of disadvantages over the dedicated loss maps. They have lower signal levels, and so a lower signal to noise ratio. This makes it hard to get clean data in lower loss regions. They contain an unquantified mix of losses from both beams and in both planes. This makes it impossible to completely separate out different sources of loss as can be done with the dedicated loss maps.

The logging database is used to identify typical data taking fills, where the squeeze was successful and the beam reached the `stable beams' mode. Then the BLM signals and optical parameters as functions of time can be retrieved for those fills. Even with no losses occurring there is a continuous low level of background noise recorded by the BLMs. We consider this noise floor to be the limit to the precision of the BLM signal and hence the uncertainty. For each BLM a background noise level is calculated by averaging the lowest 5 readings during the squeeze. This value can then be used as an estimate of the uncertainty of the BLM signal.

During a fill the rate of loss varies considerably. For much of the time the losses at the TCTs are below the noise thresholds of the BLMs. This can cause spurious values for local inefficiency. It was found that there could be large swings in the total loss rate around the fixed points of the squeeze so BLM data taken at those points are particularly unreliable.

\begin{figure}[!htbp]
	\begin{center}
		\includegraphics[width=1.0\columnwidth]{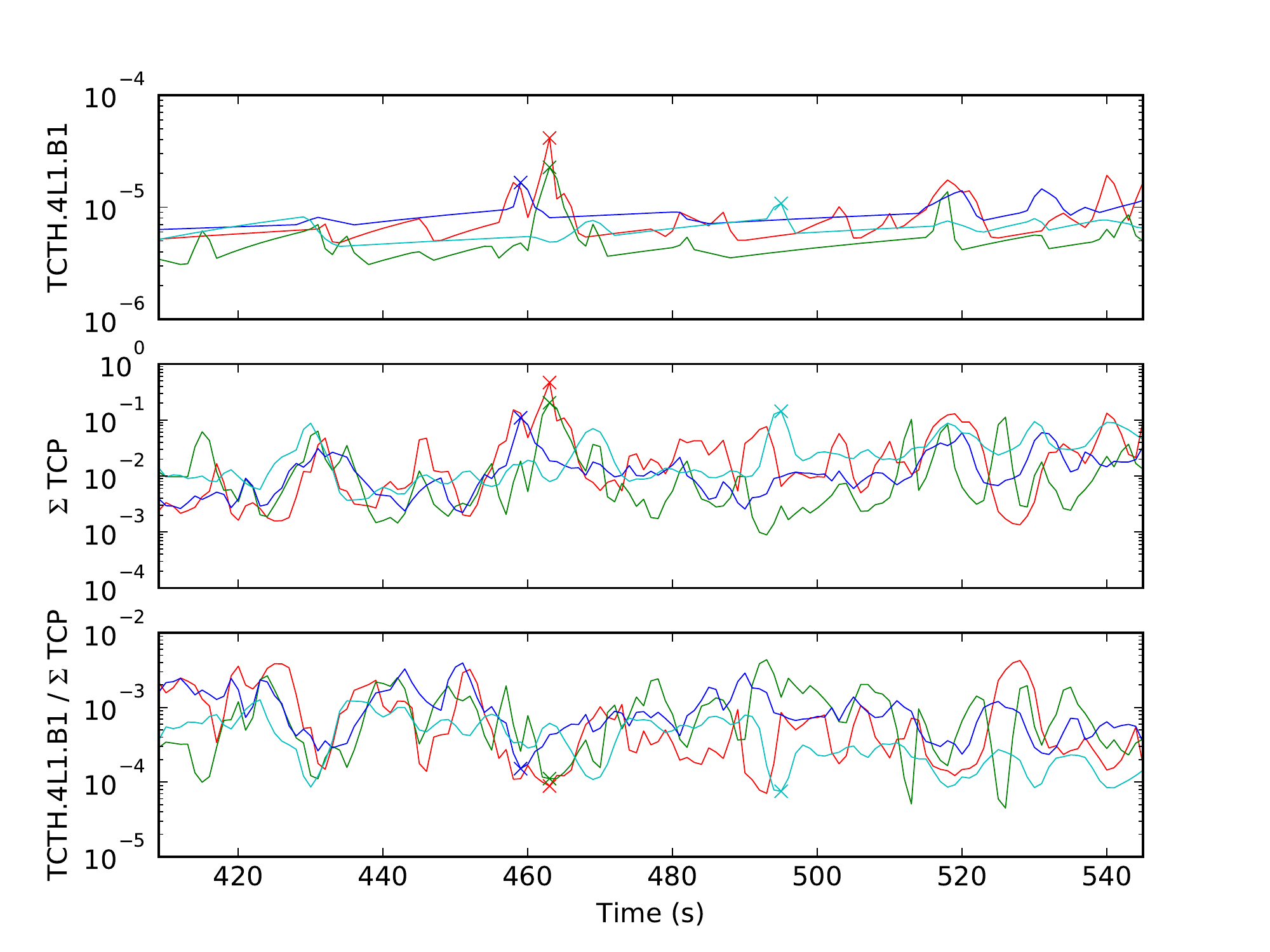}
		\caption{Peaks in the TCT (top) and TCP (middle) BLM signals are identified for 4 fills (show in different colors). At these time steps the noise on the ratio (bottom) is minimised and so can be taken to get the local cleaning inefficiency.}
		\label{fig:find_best_points_high_ir1_400}
	\end{center}
\end{figure}

In order to get a good measure of inefficiency we identified points in time where the total losses were high enough that the TCT BLMs were above noise. A peak-finding algorithm was used to find the highest values of the TCT losses within each fill. These points were retained if at the same time stamp there was also a high BLM value at TCP. Figure \ref{fig:find_best_points_high_ir1_400} shows how for four fills, time stamps with simultaneous peaks are selected. These points were then ranked by the product of the TCT and TCP values, and the highest kept.

\begin{figure}[!htbp]
	\begin{center}
		\includegraphics[width=1.0\columnwidth]{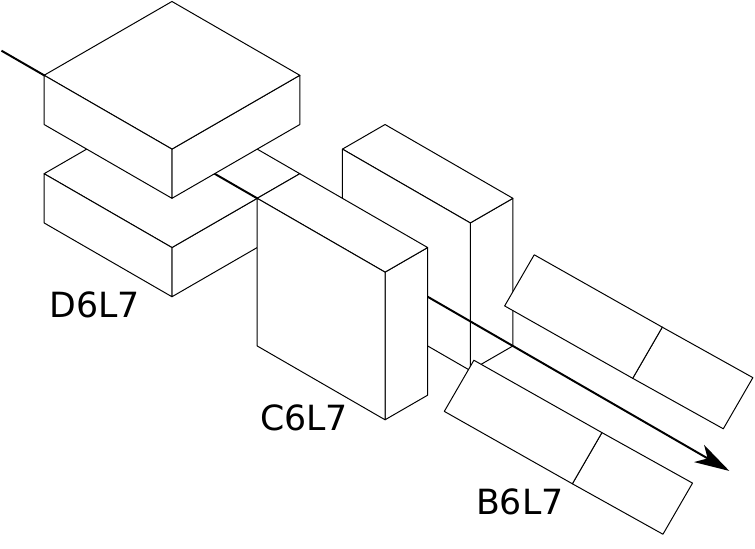}
		\caption{The 3 primary collimators at D6L7, C6L7 and B6L7 are oriented to collimate in the vertical, horizontal and skew planes respectively. The codes refer to the positions in the \engordnumber{6} cell left of IR7.}
		\label{fig:tcps}
	\end{center}
\end{figure}

The BLM signals during regular operation contain a mix of both horizontal and vertical losses. These can be partially separated by looking at the ratios between the vertical, horizontal, and skew TCPs, labelled D6L7, C6L7 and B6L7 respectively due to their positions in the lattice. A schematic of their layout is shown in Fig.~\ref{fig:tcps}. Vertical excitations will hit D6L7 leaving a BLM signal there but the shower will also leave a signal at C6L7 and B6L7. Horizontal excitations will pass through D6L7 and hit C6L7 first, with the shower peaking in B6L7. We find that cutting on the D6L7 to C6L7 ratio being less than 0.1 and the C6L7 to B6L7 ratio being less than 0.5 selects cases that are dominated by horizontal losses. This results in 13 data points that pass the filters, covering a range of \betastar values from 2~m to 0.6~m.

More sophisticated machine learning algorithms exist to categorise losses~\cite{valentino_machine_2018}, however these were not used as they have not been trained on 2012 data.

For the points that pass both filters we calculate the inefficiency at the TCTs and take the \betastar value that corresponds to the timestamp.

\subsection{Comparison between simulation and measurements}

\begin{figure}[!htbp]
	\begin{center}
		\includegraphics[width=1.0\columnwidth]{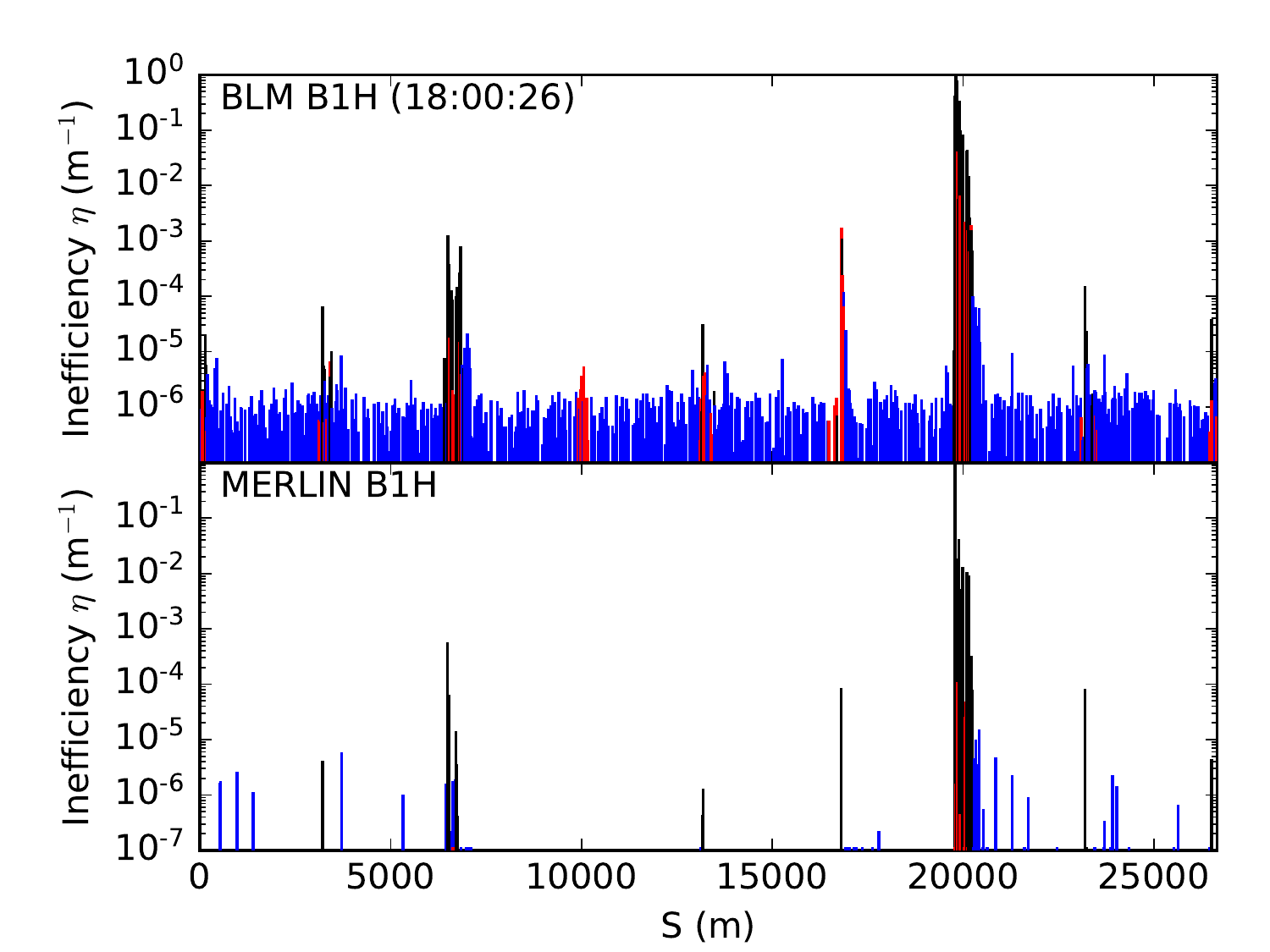}
		\caption{Beam 1 Horizontal loss map from BLMs (top) and Merlin++ (bottom) at 4 TeV with \betastar of 60~cm.}
		\label{fig:2012/Ring_B1H}
	\end{center}
\end{figure}

\begin{figure}[!htbp]
	\begin{center}
		\includegraphics[width=1.0\columnwidth]{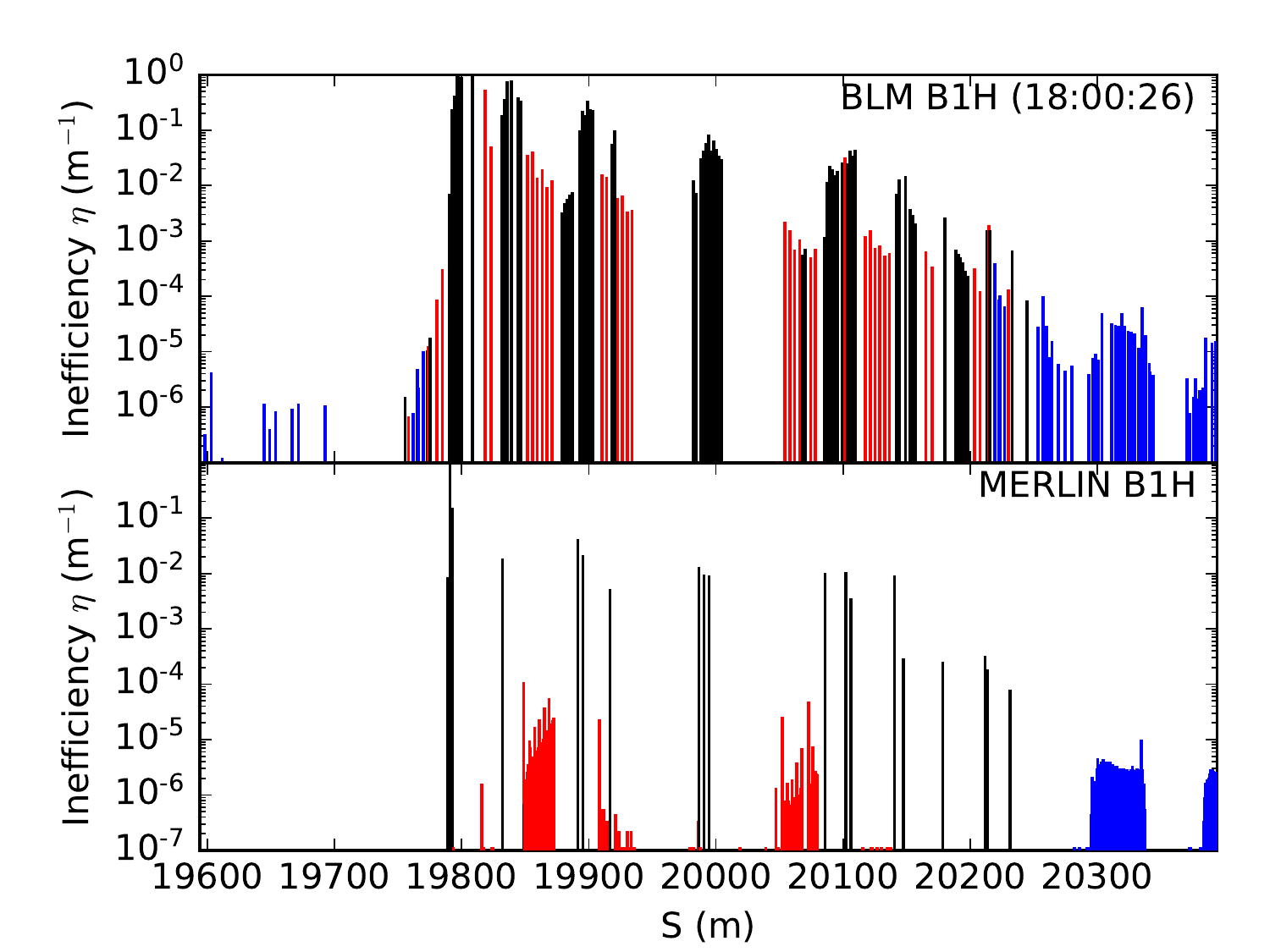}
		\caption{IR7 Beam 1 horizontal loss map from BLMs (top) and Merlin++ (bottom) at 4 TeV with \betastar of 60~cm.}
		\label{fig:2012/IR7_B1H}
	\end{center}
\end{figure}

First we can compare the full loss maps taken at the end of the squeeze with IR1/5 \betastar of 60~cm. BLM loss maps are from fill 2788 on the 1st of July 2012. Figure~\ref{fig:2012/Ring_B1H} shows losses around the full ring for horizontal beam~1 excitation. Merlin++ reproduces the significant loss peaks in the collimation regions and other IRs. Figure~\ref{fig:2012/IR7_B1H} shows the loss map zoomed to the IR7 collimation region. The hierarchy of losses from the TCPs through to the TCSs and TCLAs can bee seen in both simulation and data. The BLM signal outside the collimators is higher than in simulation, especially in the warm losses represented by the red bars, as full showers of secondary particles are not simulated in Merlin++. The noise level the BLMs can be seen in the measurements at around $10^{-6}$, these are not real losses and therefore set the precision of the measurement.

While Merlin++ counts the particle losses on the beam pipe, the BLMs record the dose from the radiation shower outside the accelerator's physical components. At LHC energies the shower from proton impacts have an effective length of approximately 1~m in typical metals with a tail expanding up to 10~m~\cite{jeanneret_quench_1996}, so proton losses at one element will also cause signal in the BLMs at downstream elements. The materials of the magnets and surrounding equipment will absorb some of the energy of the shower.
For a full quantitative comparison to the BLM signals one would need to use the proton loss maps from Merlin++ as inputs to an energy deposition code such as FLUKA. This would be used to model the evolution of the particle showers through the machine elements and the signal response of BLM ionisation chamber. Similar studies using loss maps produced with SixTrack have been demonstrated in~\cite{bruce_simulations_2014}.

We can now compare the TCT inefficiency predicted in the Merlin++ simulation with the measurements.
In both cases we are interested in the normalised local cleaning inefficiency at the TCTs, i.e. the ratio of the individual TCT to the total TCP losses. This partially normalises out the conversion of proton losses to BLM signal values. However, we must also make a normalisation to take into account the response of the BLM to the local proton loss and cross talk due to a secondary particles from one element reaching the BLM of another. To do this we normalise to the inefficiency at the fully squeezed configuration. This normalisation point is chosen as it has the highest losses and so the lowest statistical error. With the chosen normalization, the different optics configurations can be compared in relative, although not in absolute, assuming that the BLM response is independent of optics.

\begin{figure}[!htbp]
	\begin{center}
		\includegraphics[width=1.0\columnwidth]{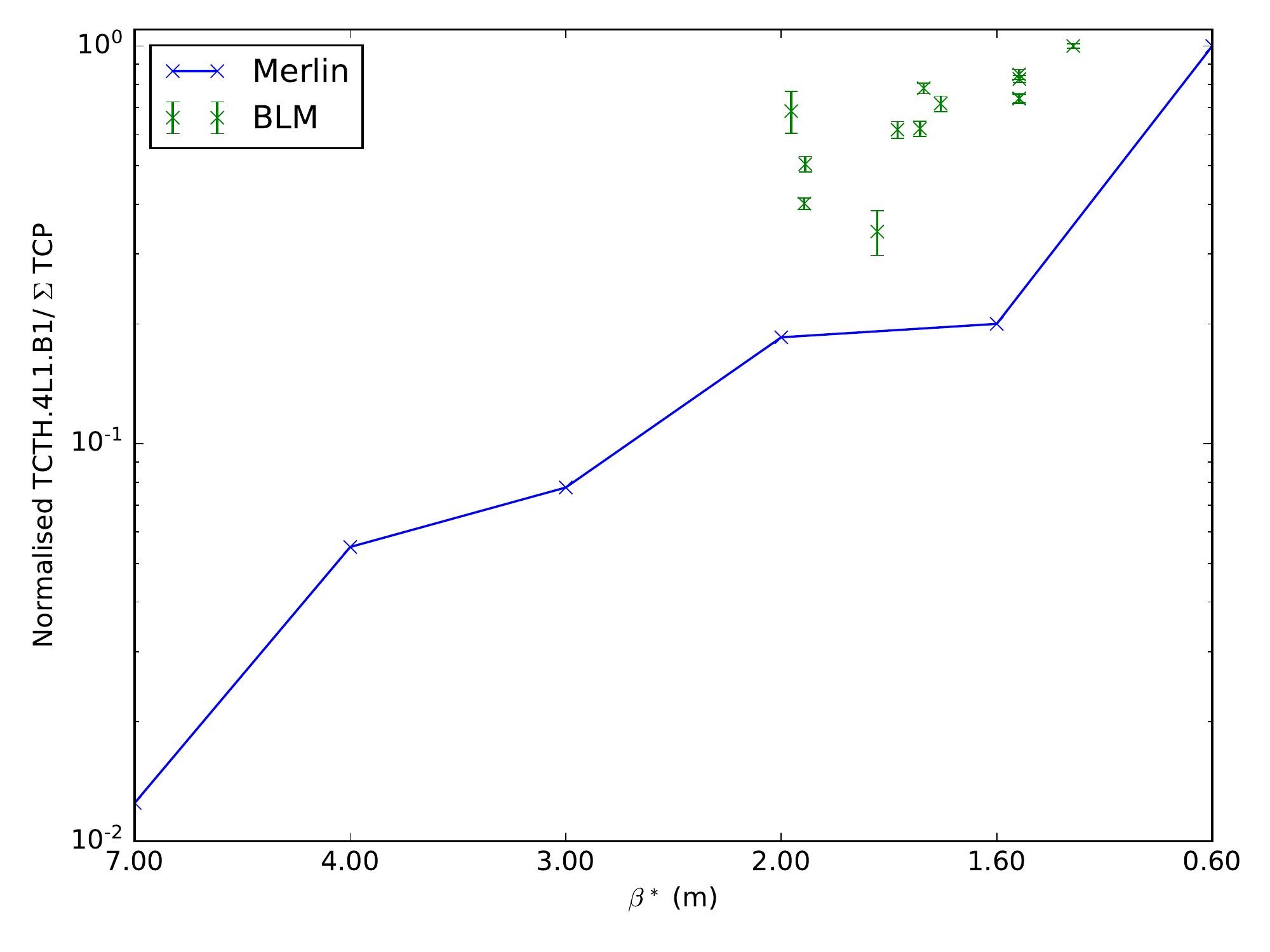}
		\caption{Comparison between normalised losses in Merlin++ and BLM signal for the horizontal collimator in IR1 (TCTH.4L1.B1) during squeeze. Note that no losses were seen at the TCT in simulation for \betastar greater than 7~m.}
		\label{fig:squeeze_loss_best_TCTH.4L1.B1}
	\end{center}
\end{figure}

Figure \ref{fig:squeeze_loss_best_TCTH.4L1.B1} shows the Merlin++ simulation compared to the data points extracted from the BLM data. As before, BLM error bars are based on the background level found by averaging the 5 lowest reading within the time window. While the trend is compatible it is clear that the BLM data are too limited to draw conclusions. The signal to noise ratio in the BLM data are too low for \betastar above 2~m to retrieve any data points and give a significant data spread above 1~m. It is clear that dedicated loss maps are needed to make a better comparison.

\section{Run 2 - 6.5 TeV}
\label{sec:run2}

\subsection{Run 2 2016 LHC configuration}

Run 2 of the LHC began in 2015 and incorporated changes to the machine configuration, most notably an increase in beam energy from 4 to 6.5~TeV.
In order to reduce the time from injection to collision, a combined ramp and squeeze program was used stating from 2016, such that the initial squeeze, down to \betastar of 3~m at IR1 and IR5, happens simultaneously with the energy ramp. 
Therefore, the squeeze beam mode covers just the final 3~m to 0.4~m of squeezing.

Table \ref{tab:optics2016} shows the optical parameters for the IPs used during 2016 data taking. Table \ref{tab:collimators2016} shows the collimation settings used.

\begin{table}[!ht]
\caption{Optics settings for 2016 squeeze at 6.5~TeV. As before for IR2 and 8 the external crossing angle is given.}
\centering
\begin{tabular}{|l|l|l|l|l|}
\hline
  & \betastar &  half crossing angle \\
  &        (m)& (\si{\micro rad})\\
\hline
ATLAS (IP1)&  3.0 $\rightarrow$ 0.4& -185 V\\
CMS   (IP5)&  3.0 $\rightarrow$ 0.4& 185 H\\
ALICE (IP2)&  10 $\rightarrow$ 10& 200 V\\
LHCb  (IP8)&  6.0 $\rightarrow$ 3.0& -250 H\\
\hline
\end{tabular}
\label{tab:optics2016}
\end{table}

\begin{table}[!ht]
\caption{Collimator settings for the squeeze in 2016 at 6.5~TeV. Using a normalised beam emittance of 3.5~\si{\micro m}}
\centering
\begin{tabular}{|l|l|l|}
\hline
Region & Type & Gap ($\upsigma$)\\
\hline
IR7 & TCP & 5.5\\
IR7 & TCS & 7.5\\
IR7 & TCLA & 11.0\\
\hline
IR3 & TCP & 15.0\\
IR3 & TCS & 18.0\\
IR3 & TCLA & 20.0\\
\hline
IR1 & TCT & 23.0 $\rightarrow$ 9.0\\
IR5 & TCT & 23.0 $\rightarrow$ 9.0\\
IR2 & TCT & 37.0\\
IR8 & TCT & 23.0 $\rightarrow$ 15.0\\
\hline
\end{tabular}
\label{tab:collimators2016}
\end{table}

\subsection{Measured squeeze losses in 2016}

\begin{figure}[!htbp]
	\begin{center}
		\includegraphics[width=1.0\columnwidth]{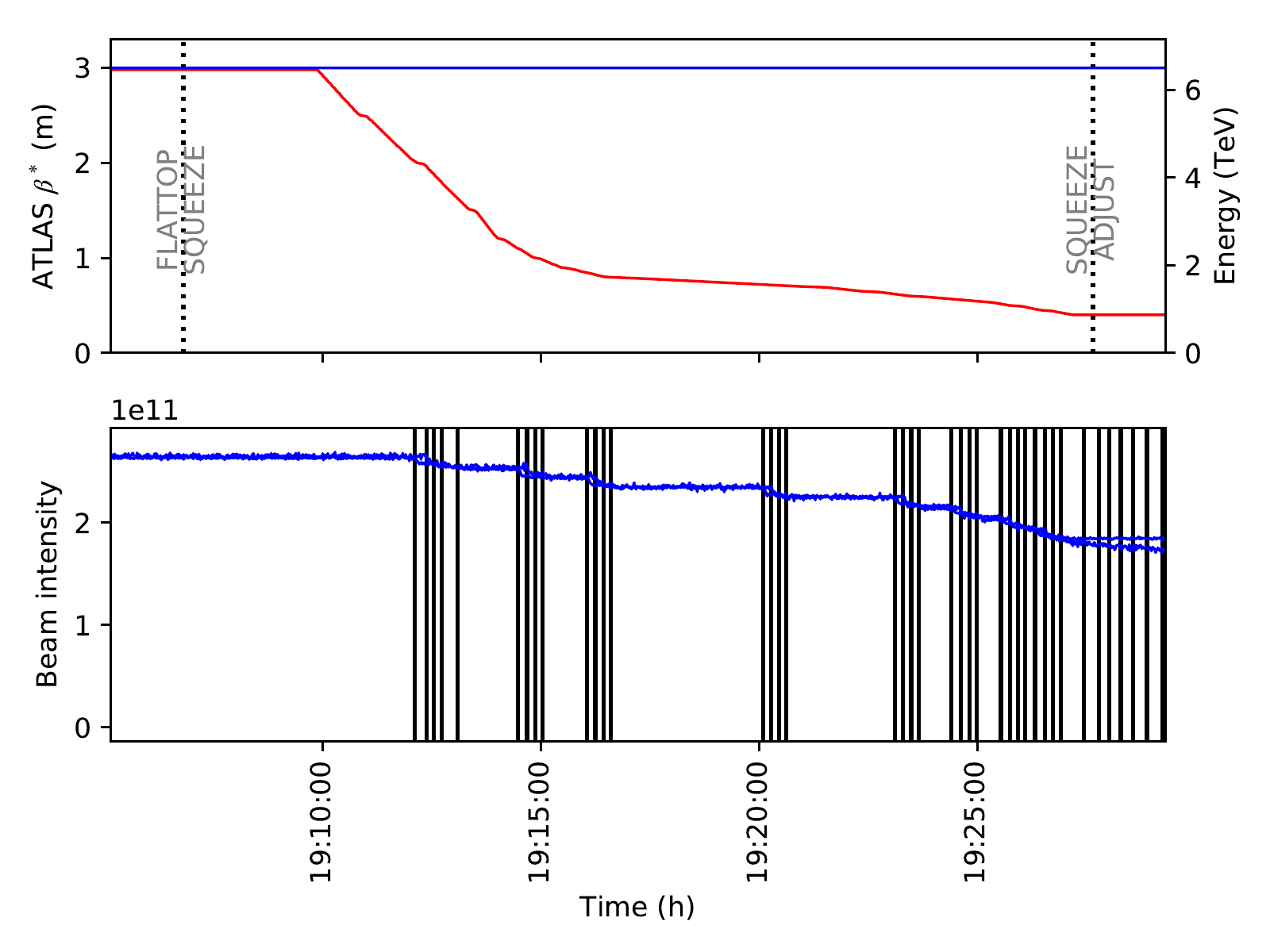}
		\caption{Loss map recording during fill 4832. Top plot shows energy (blue) and IR1 \betastar (red). Bottom plot shows the fall in beam intensity as each of the 4 ADTs (for each beam and plane) are fired (vertical lines).}
		\label{fig:fill_4832}
	\end{center}
\end{figure}

During the 2016 beam commissioning a number of loss maps were made during the squeeze. This gives a better signal to noise ratio and allows separation of losses from each beam and plane. The maps used in this article were taken on 20th of April 2016, during fill 4832. A beam of low intensity pilot bunches was injected and ramped to 6.5~TeV. During the squeeze, the ADTs for each combination of horizontal and vertical, and Beam 1 and 2, were fired in turn to excite one of the bunches in that plane, as shown in Fig.~\ref{fig:fill_4832}, and the BLM signal was recorded~\cite{mirarchi_collimation:_2016}. 

For each BLM we calculate a background level, by averaging the signal during a 10~s window near the start of the squeeze where losses are low. This fixed value per BLM is used as an estimate of the uncertainty of that BLM's signal during excitation. Note that the background measurement is usually taken closer to the loss map excitation, however in this case where loss maps are made in rapid succession this is not possible. There are a number of additional parameters not under control that can contribute to errors, such as orbit shifts and changes in the squeeze rate.

\subsection{Comparison between simulation and measurements}

\begin{figure}[!htbp]
	\begin{center}
		\includegraphics[width=1.0\columnwidth]{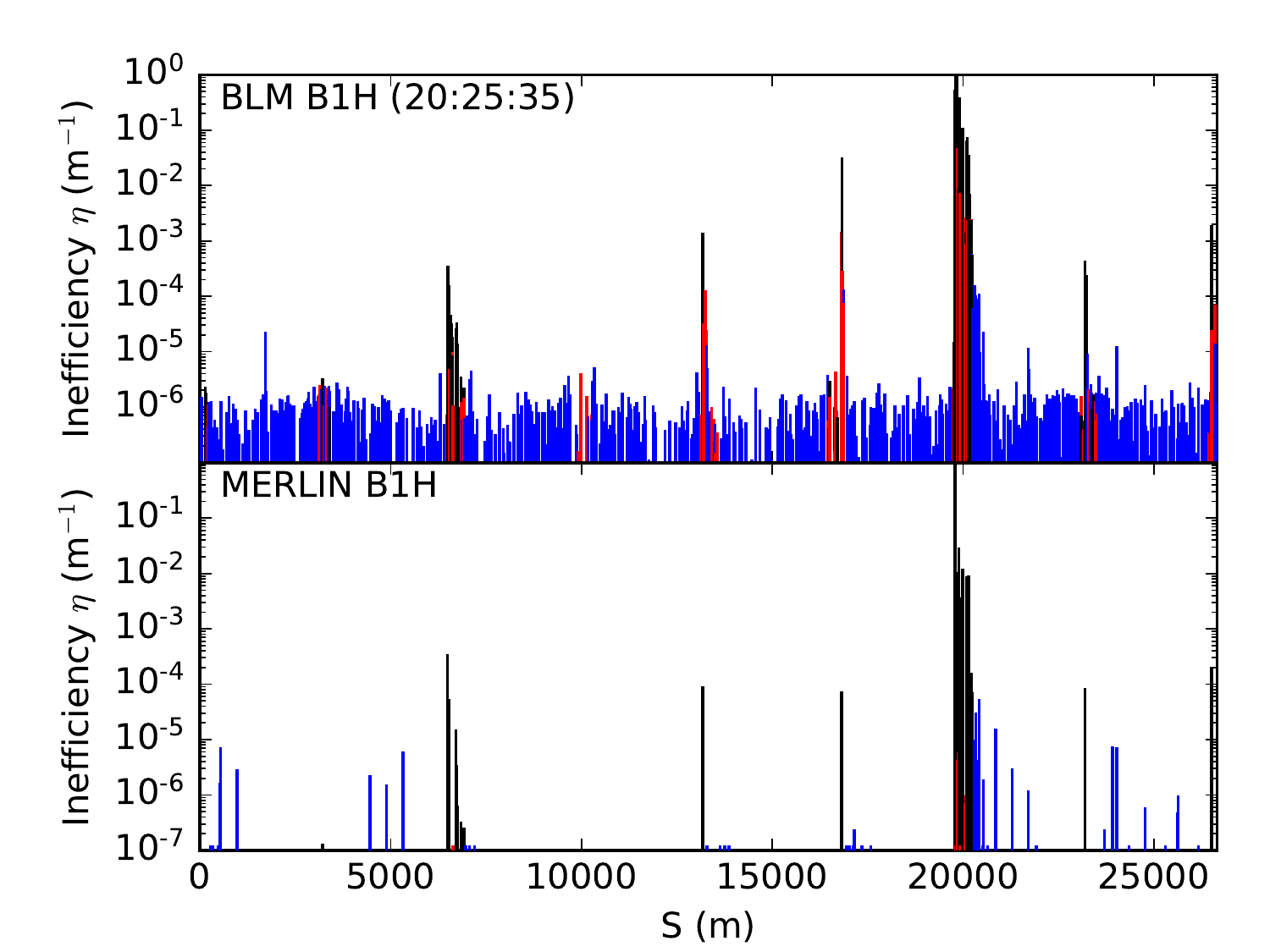}
		\caption{Beam 1 Horizontal loss map from BLMs (top) and Merlin++ (bottom) at 6.5 TeV with \betastar of 50~cm.}
		\label{fig:2016/Ring_B1H}
	\end{center}
\end{figure}

\begin{figure}[!htbp]
	\begin{center}
		\includegraphics[width=1.0\columnwidth]{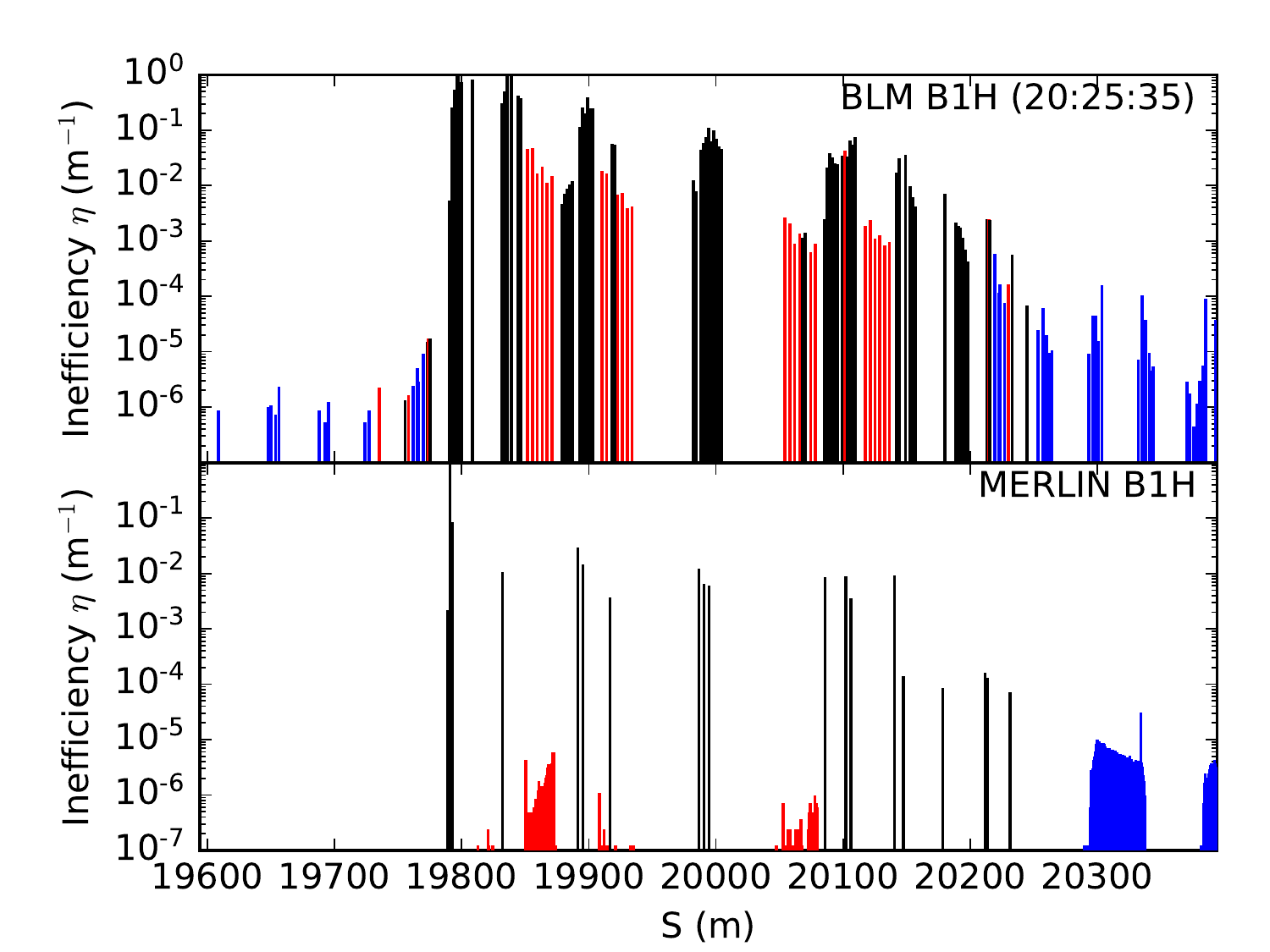}
		\caption{IR7 Beam 1 horizontal loss map from BLMs (top) and Merlin++ (bottom) at 6.5 TeV with \betastar of 50~cm.}
		\label{fig:2016/IR7_B1H}
	\end{center}
\end{figure}

First we compare a full loss map from fill 4832 taken close to when the \betastar at IR1/5 crossed 50~cm. Figures \ref{fig:2016/Ring_B1H} and \ref{fig:2016/IR7_B1H} show full ring and IR7 loss maps comparing BLM data and Merlin++ simulation. As with the 4~TeV comparisons we see that Merlin++ reproduces well the main loss locations around the ring, and the collimation hierarchy in IR7 well.

We can now compare the normalised cleaning efficiency as function of \betastar between BLM data and Merlin++. In the BLM measurements during the squeeze we have 11 loss maps in the horizontal plane and 9 in the vertical. In the simulations we have used 5 different optical configurations.

\begin{figure}[!htbp]
	\begin{center}
		\includegraphics[width=1.0\columnwidth]{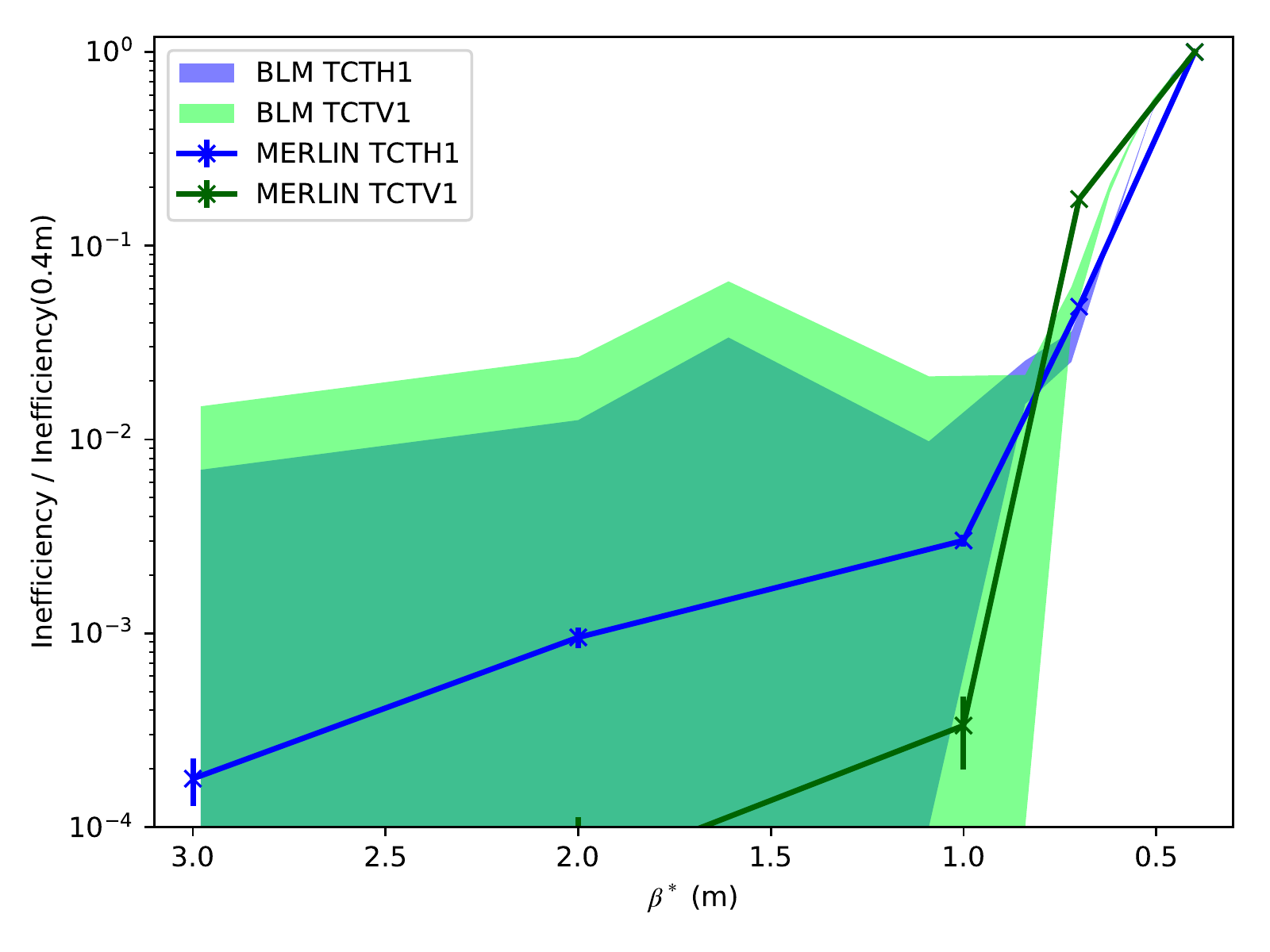}
		\caption{BLM signal (shown as uncertainty bands) and Merlin++ simulated losses (shown as solid lines) on IR1 TCTs for horizontal excitation. Solid area shows uncertainty bands due to detector background.}
		\label{fig:merlin_norm_6p5tev_b1_h_ir1}
	\end{center}
\end{figure}

\begin{figure}[!htbp]
	\begin{center}
		\includegraphics[width=1.0\columnwidth]{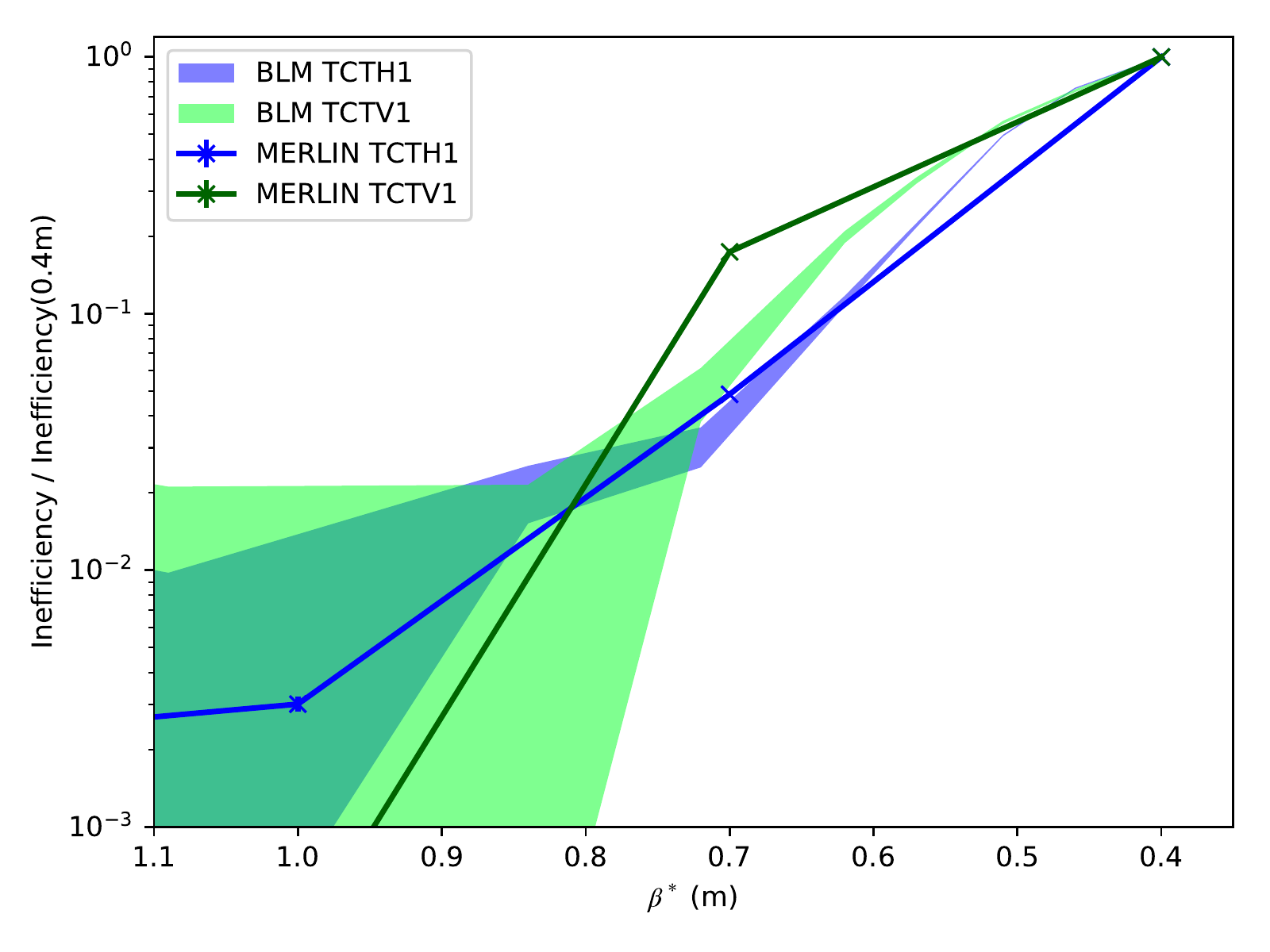}
		\caption{Zoomed plot of BLM signal and Merlin++ simulated losses on IR1 TCTs for horizontal excitation. Solid area shows uncertainty bands due to detector background.}
		\label{fig:merlin_norm_6p5tev_b1_h_ir1_zoom}
	\end{center}
\end{figure}

\begin{figure}[!htbp]
	\begin{center}
		\includegraphics[width=1.0\columnwidth]{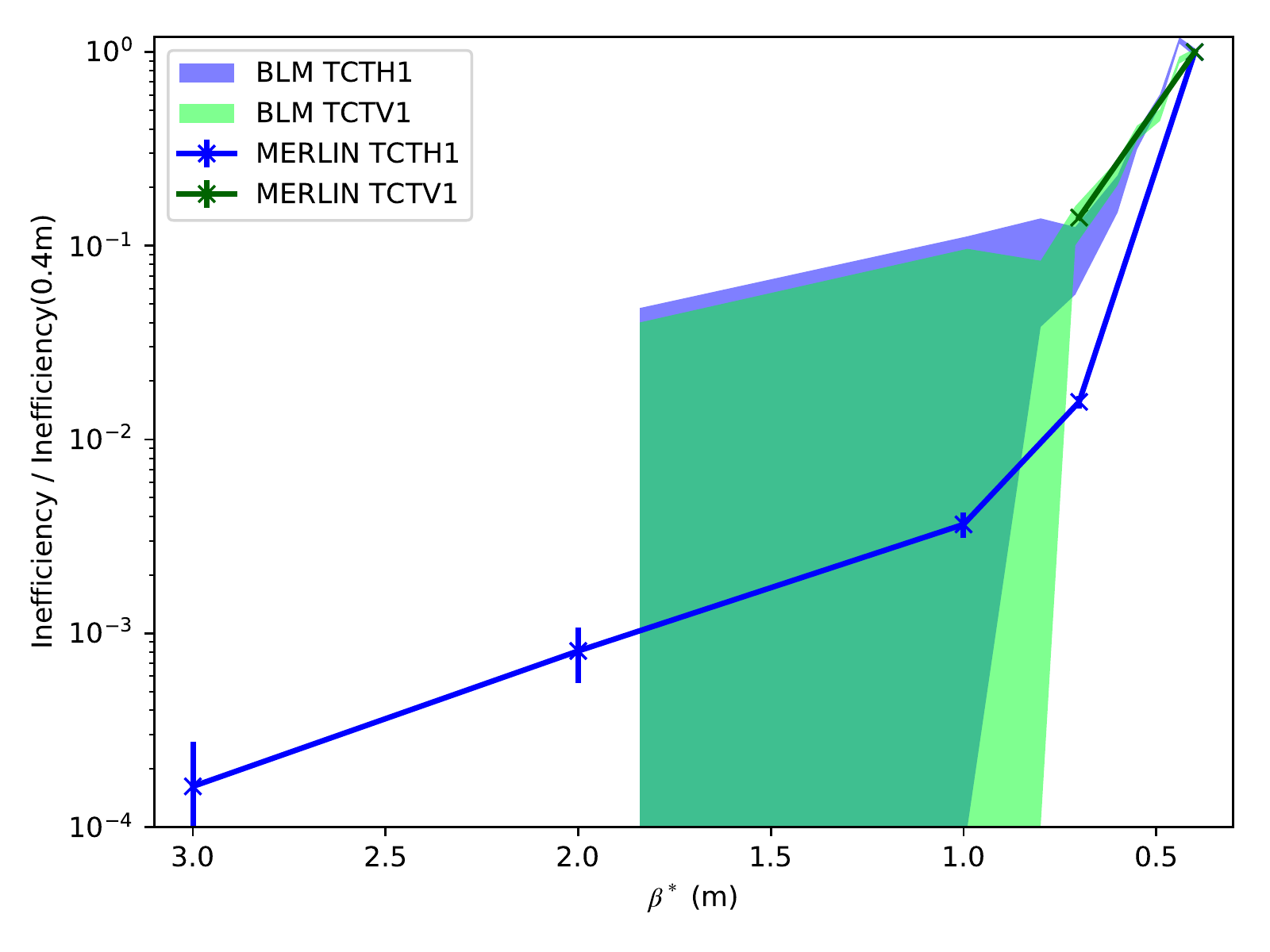}
		\caption{BLM signal and Merlin++ simulated losses on IR1 TCTs for vertical excitation. Note that no losses were seen on the vertical TCT at \betastar greater than 0.8~m in simulation or on either TCT at \betastar greater than 1.9~m in data.}
		\label{fig:merlin_norm_6p5tev_b1_v_ir1}
	\end{center}
\end{figure}

\begin{figure}[!htbp]
	\begin{center}
		\includegraphics[width=1.0\columnwidth]{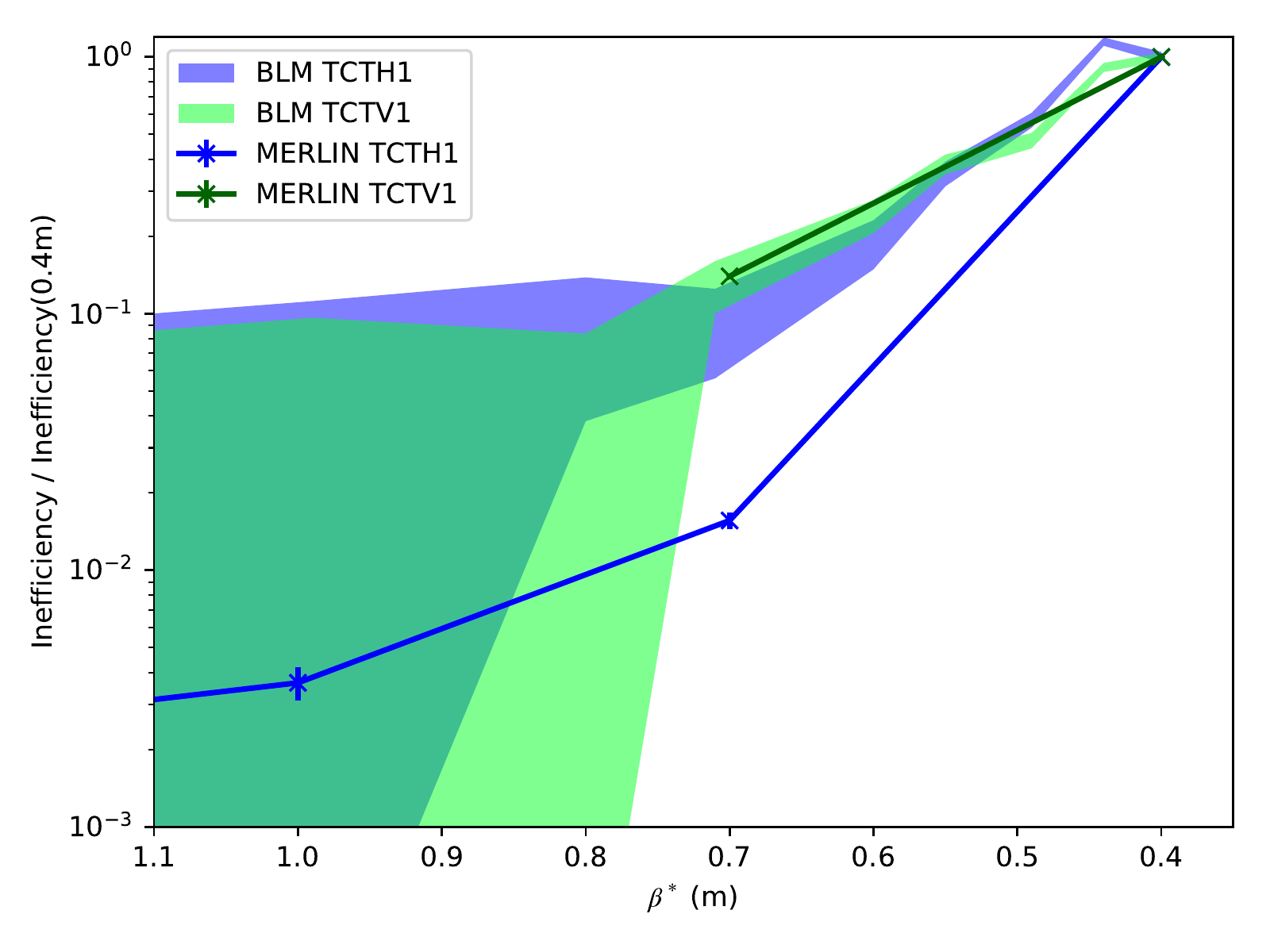}
		\caption{Zoomed plot of BLM signal and Merlin++ simulated losses on IR1 TCTs for vertical excitation.}
		\label{fig:merlin_norm_6p5tev_b1_v_ir1_zoom}
	\end{center}
\end{figure}

Figures \ref{fig:merlin_norm_6p5tev_b1_h_ir1} and \ref{fig:merlin_norm_6p5tev_b1_v_ir1} show losses on the IR1 TCTs during the squeeze due to horizontal and vertical excitation of the beam. Figures \ref{fig:merlin_norm_6p5tev_b1_h_ir1_zoom} and \ref{fig:merlin_norm_6p5tev_b1_v_ir1_zoom} show zoomed sections of the plot so that more detail is visible at low \betastar.  Horizontally we see excellent agreement between data and simulation, with steep increases in TCT losses as the beam is squeezed to \betastar of 0.4~m. For vertical excitation we again see good overall agreement, although no losses are observed on TCTPV.4L1.B1, the vertical TCT in IR1, in simulation above \betastar of 0.8~m. At larger \betastar values, the signals on the TCT BLMs are below the noise levels, so we are not able to record the losses.



\begin{figure}[!htbp]
	\begin{center}
		\includegraphics[width=1.0\columnwidth]{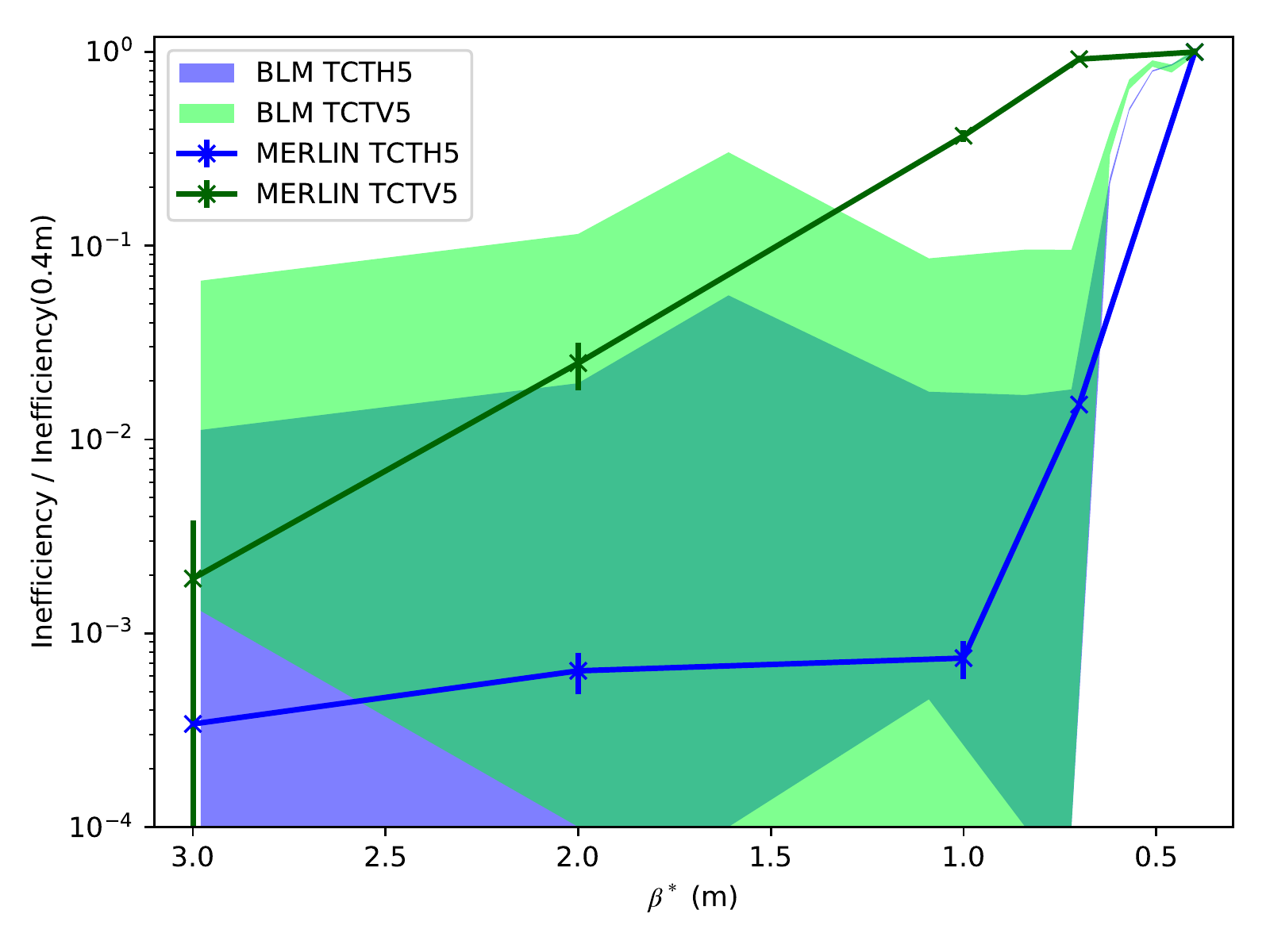}
		\caption{BLM signal and Merlin++ simulated losses on IR5 TCTs for horizontal excitation.}
		\label{fig:merlin_norm_6p5tev_b1_h_ir5}
	\end{center}
\end{figure}

\begin{figure}[!htbp]
	\begin{center}
		\includegraphics[width=1.0\columnwidth]{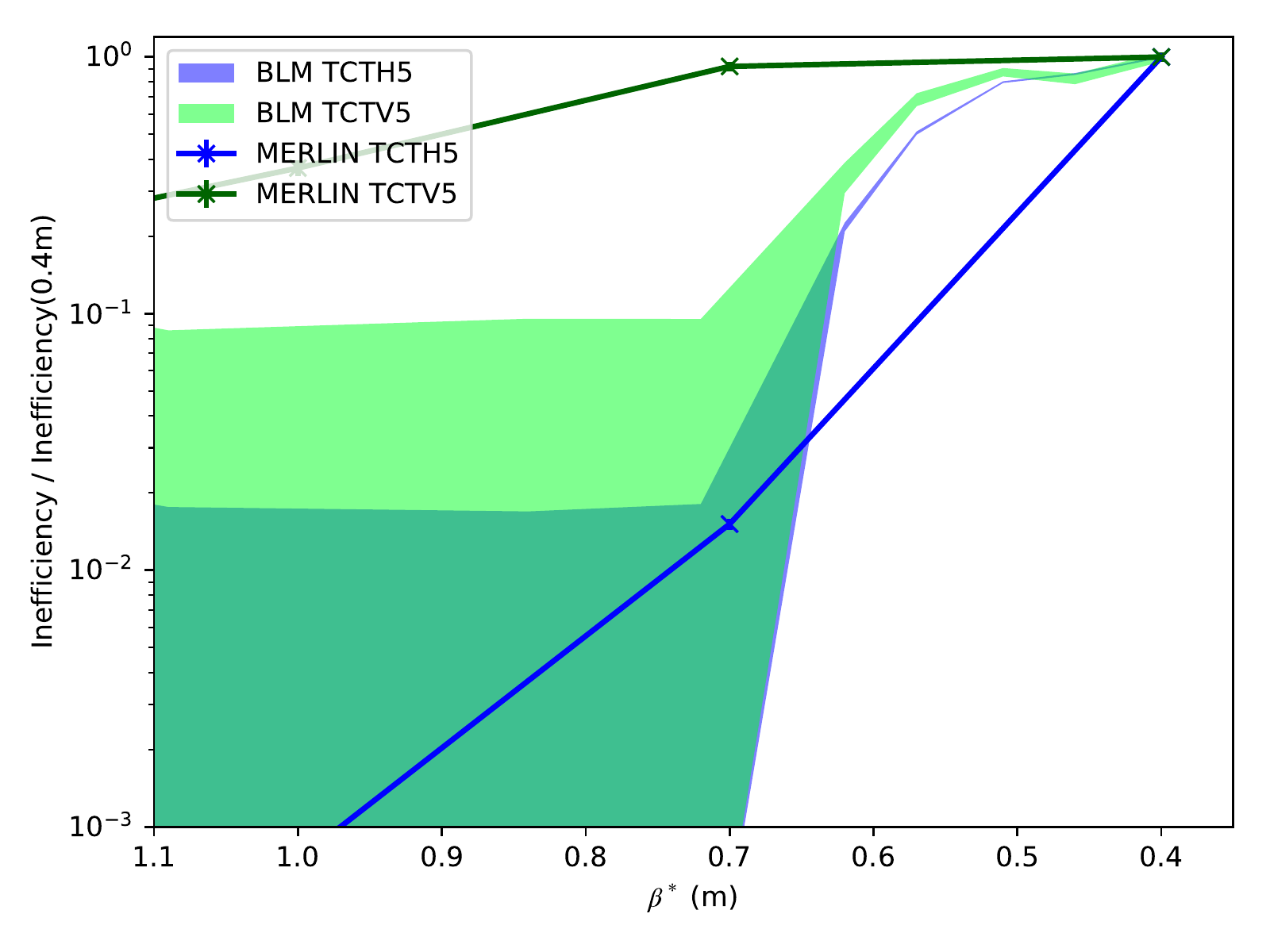}
		\caption{Zoomed plot of BLM signal and Merlin++ simulated losses on IR5 TCTs for horizontal excitation.}
		\label{fig:merlin_norm_6p5tev_b1_h_ir5_zoom}
	\end{center}
\end{figure}

\begin{figure}[!htbp]
	\begin{center}
		\includegraphics[width=1.0\columnwidth]{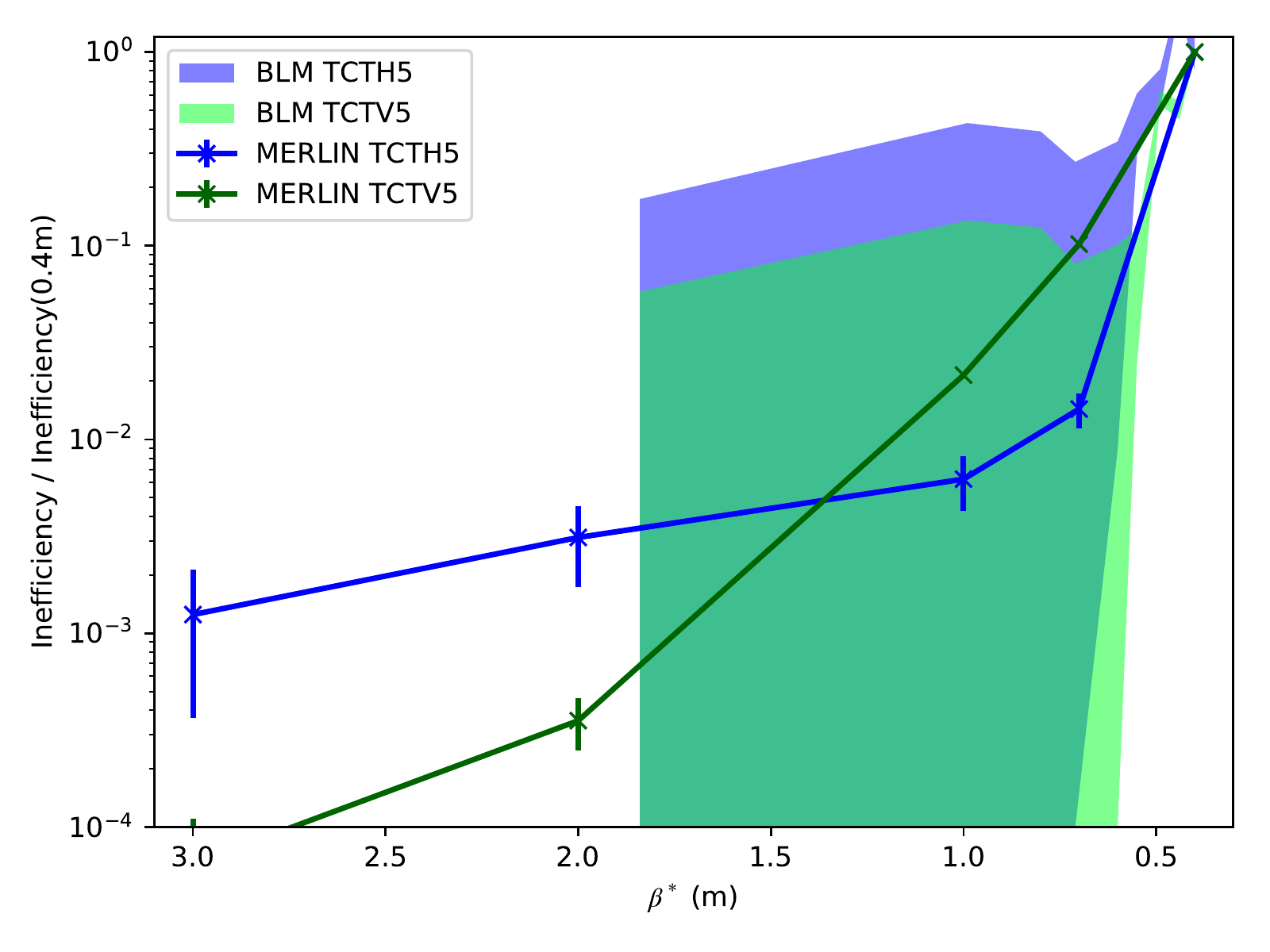}
		\caption{BLM signal and Merlin++ simulated losses on IR5 TCTs for vertical excitation.}
		\label{fig:merlin_norm_6p5tev_b1_v_ir5}
	\end{center}
\end{figure}

\begin{figure}[!htbp]
	\begin{center}
		\includegraphics[width=1.0\columnwidth]{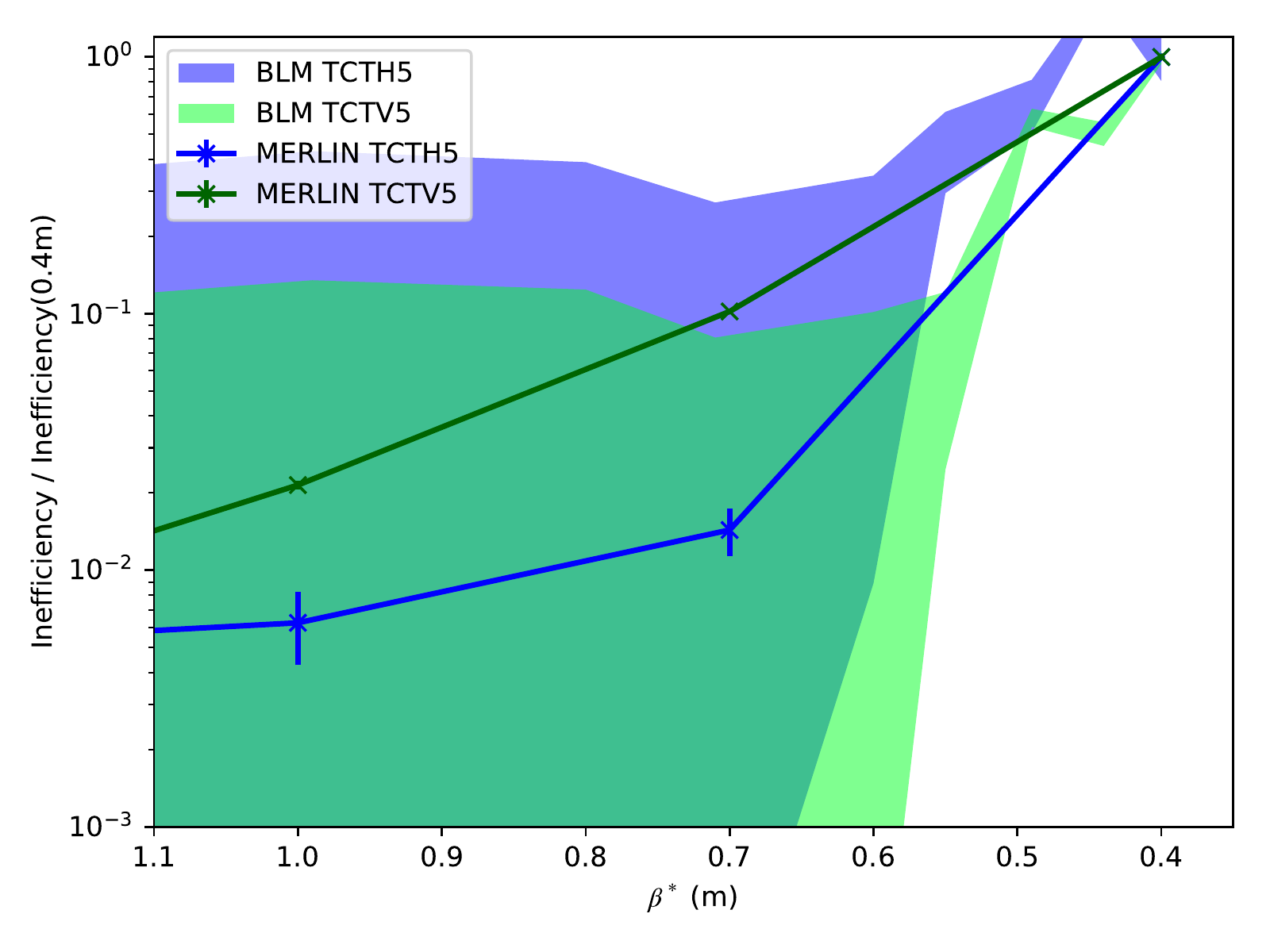}
		\caption{Zoomed plot of BLM signal and Merlin++ simulated losses on IR5 TCTs for vertical excitation.}
		\label{fig:merlin_norm_6p5tev_b1_v_ir5_zoom}
	\end{center}
\end{figure}

Figures \ref{fig:merlin_norm_6p5tev_b1_h_ir5} and \ref{fig:merlin_norm_6p5tev_b1_v_ir5} show losses on the IR5 TCTs due to horizontal and vertical excitation of the beam. Again we show zoomed sections for low \betastar in Figs. \ref{fig:merlin_norm_6p5tev_b1_h_ir5_zoom} and \ref{fig:merlin_norm_6p5tev_b1_v_ir5_zoom}. For horizontal excitation we see good agreement for TCTPH.4L5.B1, but higher losses in simulation for TCTPV.4L5.B1 than in BLM data. For vertical excitation Merlin++ reproduces the losses well.



\begin{figure}[htbp]
    \centering
    \begin{subfigure}[b]{0.49\columnwidth}
        \includegraphics[width=\columnwidth]{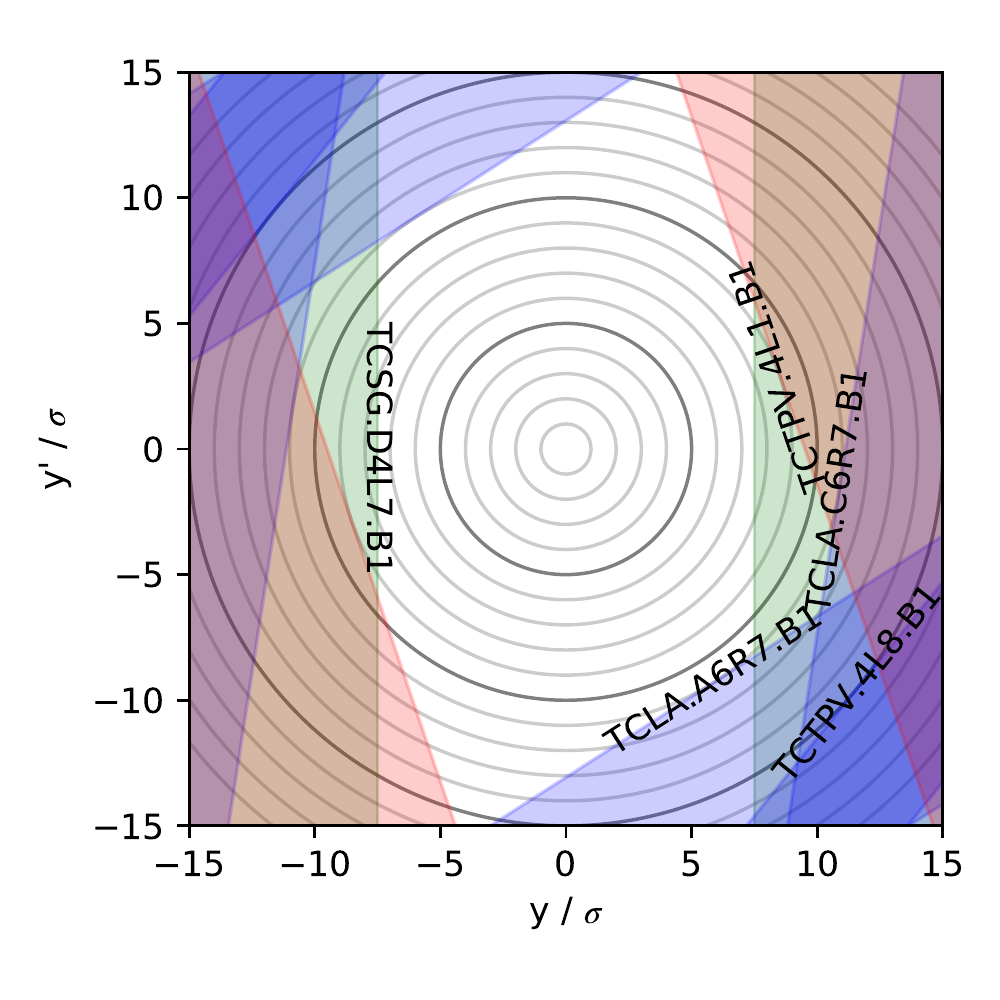}
        \caption{\betastar = 40~cm}
        \label{fig:col_phase_advance_v_TCSG.D4L7.B1_TCTPV.4L1.B1_0.4}
    \end{subfigure}
    \begin{subfigure}[b]{0.49\columnwidth}
        \includegraphics[width=\columnwidth]{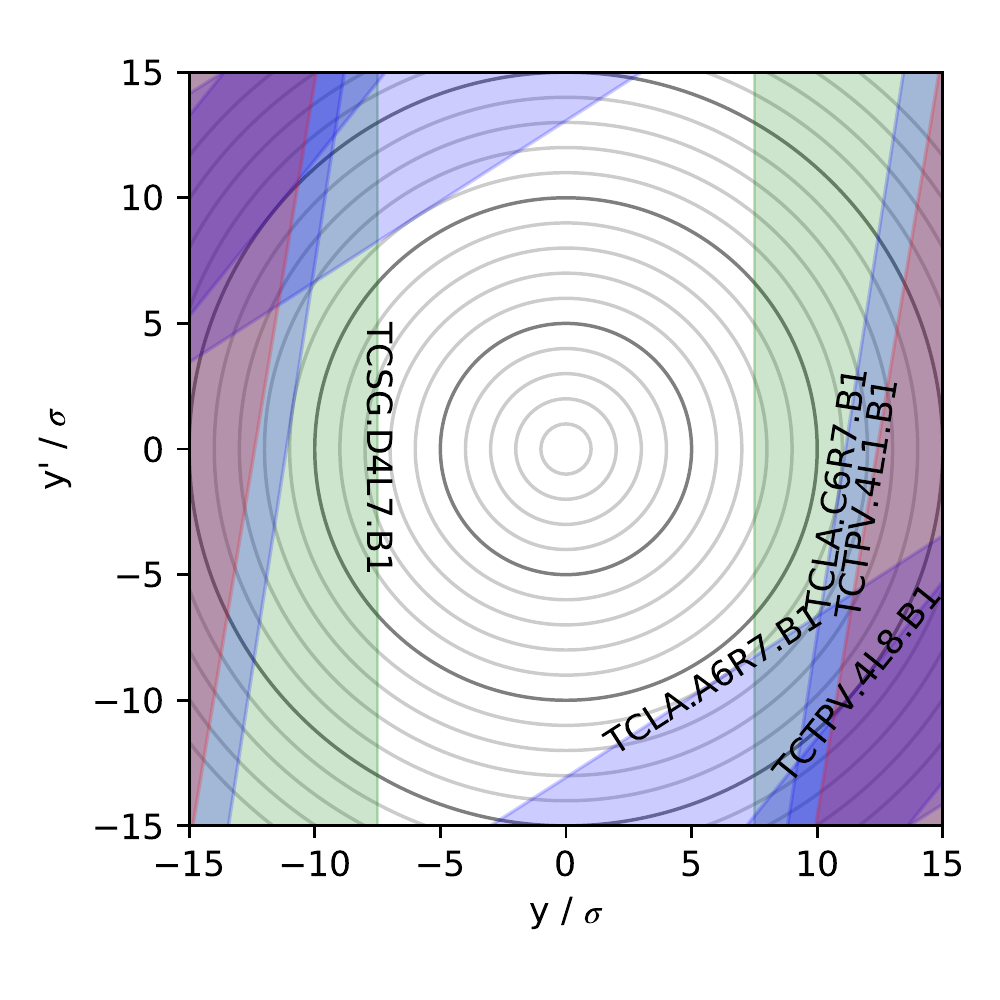}
        \caption{\betastar = 1~m}
        \label{fig:col_phase_advance_v_TCSG.D4L7.B1_TCTPV.4L1.B1_1.0}
    \end{subfigure}
    \caption{Collimator positions in normalised phase-space, accounting for the phase advance from TCSG.D4L7.B1, shown in green, to TCTPV.4L1.B1 shown in red.}
    \label{fig:col_phase_advance_v_TCSG.D4L7.B1_TCTPV.4L1.B1}
\end{figure}

To investigate the loss fall off on TCTPV.4L1.B1 (Fig.~\ref{fig:merlin_norm_6p5tev_b1_v_ir1}) we look at the position of the collimators that act in the vertical plane, projected into phase-space. At the smallest \betastar values, losses on TCTPV.4L1.B1 are dominated by particles scattered from the IR7 TCSGs with the highest losses coming from TCSG.D4L7.B1. Figure \ref{fig:col_phase_advance_v_TCSG.D4L7.B1_TCTPV.4L1.B1} shows the vertical collimators with their phase advance from TCSG.D4L7.B1. It can be seen that due to the retraction in jaw gap and change in phase advance the TCT is shadowed behind TCLA.C6R7.B1 for \betastar of 1~m and larger. In the LHC TCTPV.4L1.B1 is positioned downstream of TCTPH.4L1.B1, so the BLM will see local showers from the horizontal TCT even when the vertical TCT is not directly hit. This explains the discrepancy between the simulation and BLM data.

With this good modelling of proton losses we can now use Merlin++ to make predictions for future collimation configuration such as the HL-LHC.

\section{Performance of the HL-LHC collimation system}
\label{sec:hl-lhc}

The HL-LHC upgrade introduces several changes to the lattice~\cite{apollinari_high-luminosity_2017}. Among other changes, the inner triplets are replaced with higher gradient magnets of larger aperture to allow a smaller \betastar at the IPs~\cite{todesco_superconducting_2015}, and a new achromatic telescopic squeeze (ATS) optics scheme is used~\cite{fartoukh_achromatic_2013}. In the dispersion matching section downstream of the betatron cleaning, additional absorbers (TCLD) have been placed by splitting two of the bending magnets into shorter high field magnets~\cite{bruce_cleaning_2014, lechner_power_2014, mirarchi_cleaning_2016}. For each beam in cell 6 upstream of the experiment in IR1 and IR5 an additional pair of TCTs has been placed to improve protection \cite{redaelli_cleaning_2015}.

\begin{table}[!ht]
\caption{Collimator settings for HL-LHC at 7~TeV squeeze from 45~cm to 15~cm. Using beam emittance of 3.5~\si{\micro m}.}
\centering
\begin{tabular}{|l|l|l|}
\hline
Region & Type & Gap ($\upsigma$)\\
\hline
IR7 & TCP & 5.7\\
IR7 & TCS & 7.7\\
IR7 & TCLD & 12.0\\
\hline
IR3 & TCP & 15.0\\
IR3 & TCS & 18.0\\
\hline
IR1 & TCT & 18.2 $\rightarrow$ 10.5\\
IR5 & TCT & 18.2 $\rightarrow$ 10.5\\
IR2 & TCT & 30.0\\
IR8 & TCT & 30.0\\
\hline
\end{tabular}
\label{tab:collimators_hllhc}
\end{table}

\subsection{HL-LHC luminosity levelling}

In order to maximise integrated luminosity while limiting the maximum pileup, the HL-LHC will use a luminosity levelling scheme~\cite{arduini_implications_2015}. If the accelerator configuration is kept constant during data taking then the luminosity will fall over the length of the fill due to the gradual reduction in the beam current. Levelling is achieved by adjusting the machine configuration to compensate for the change in beam current; in the baseline by changing \betastar at the IPs.

This leads to another situation where dynamic changes of the collimators could be needed, although in this case with the beams in collision.

We consider a levelling scheme that utilises changes in \betastar from 64~cm to 15~cm, while keeping the crossing angle fixed~\cite{metral_update_2018}. In this case the TCTs and TCLs are held a fixed position in mm. The jaws are fixed at the position that gives a TCT gap of 10.5~$\sigma$ and a TCL gap of 12~$\sigma$ at the minimum \betastar of 15~cm, using a normalised beam emittance of 3.5~\si{\micro m}. For example TCTPH.4L1.B1 will have a gap of 15.5~mm for all \betastar values. Table~\ref{tab:collimators_hllhc} shows the collimator settings used.
For this work we use the HL-LHC version 1.2 optics, with 2 TCLDs per beam in IR7.

\begin{figure*}[!htbp]
	\begin{center}
		\includegraphics[width=0.75\textwidth]{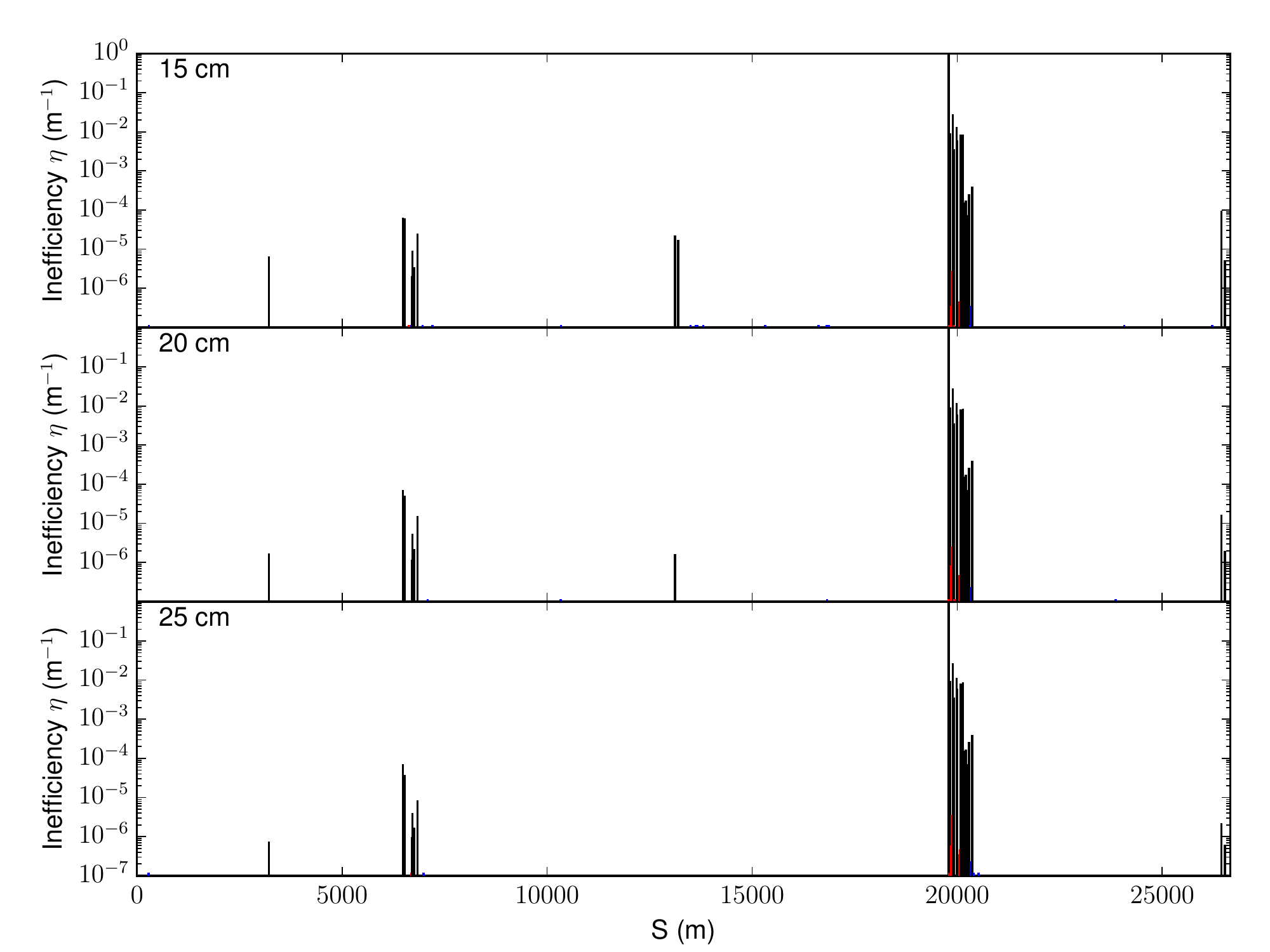}
		\caption{Merlin++ Beam 1 loss map for 3 IR1/5 \betastar value steps during HL-LHC luminosity levelling.}
		\label{fig:Ring_h_hl_level}
	\end{center}
\end{figure*}

\begin{figure*}[!htbp]
	\begin{center}
		\includegraphics[width=0.75\textwidth]{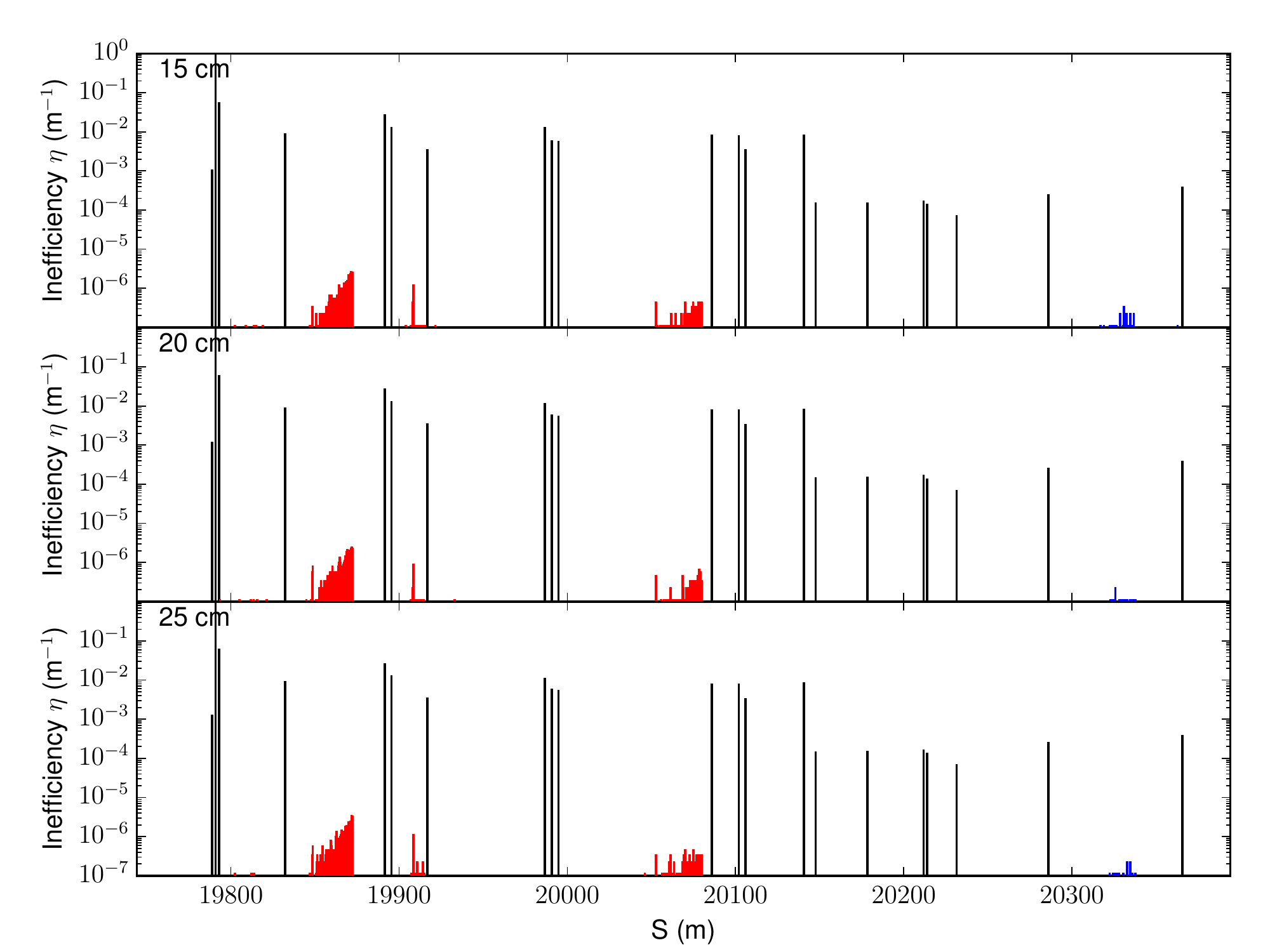}
		\caption{Merlin++ Beam 1 loss map for 3 IR1/5 \betastar value steps during HL-LHC luminosity levelling. For IR7.}
		\label{fig:IR7_h_hl_level}
	\end{center}
\end{figure*}

Figure~\ref{fig:Ring_h_hl_level} and \ref{fig:IR7_h_hl_level} show the simulated loss map at 3 steps in the HL-LHC luminosity levelling for the full ring and IR7 respectively. Again the main losses occur in the collimation regions at IR 3 and 7. Smaller loss peaks can also be seen at IR 1,2 and 5. The TCT losses get larger as \betastar at the IPs is reduced.

\begin{figure}[!htbp]
	\begin{center}
		\includegraphics[width=1.0\columnwidth]{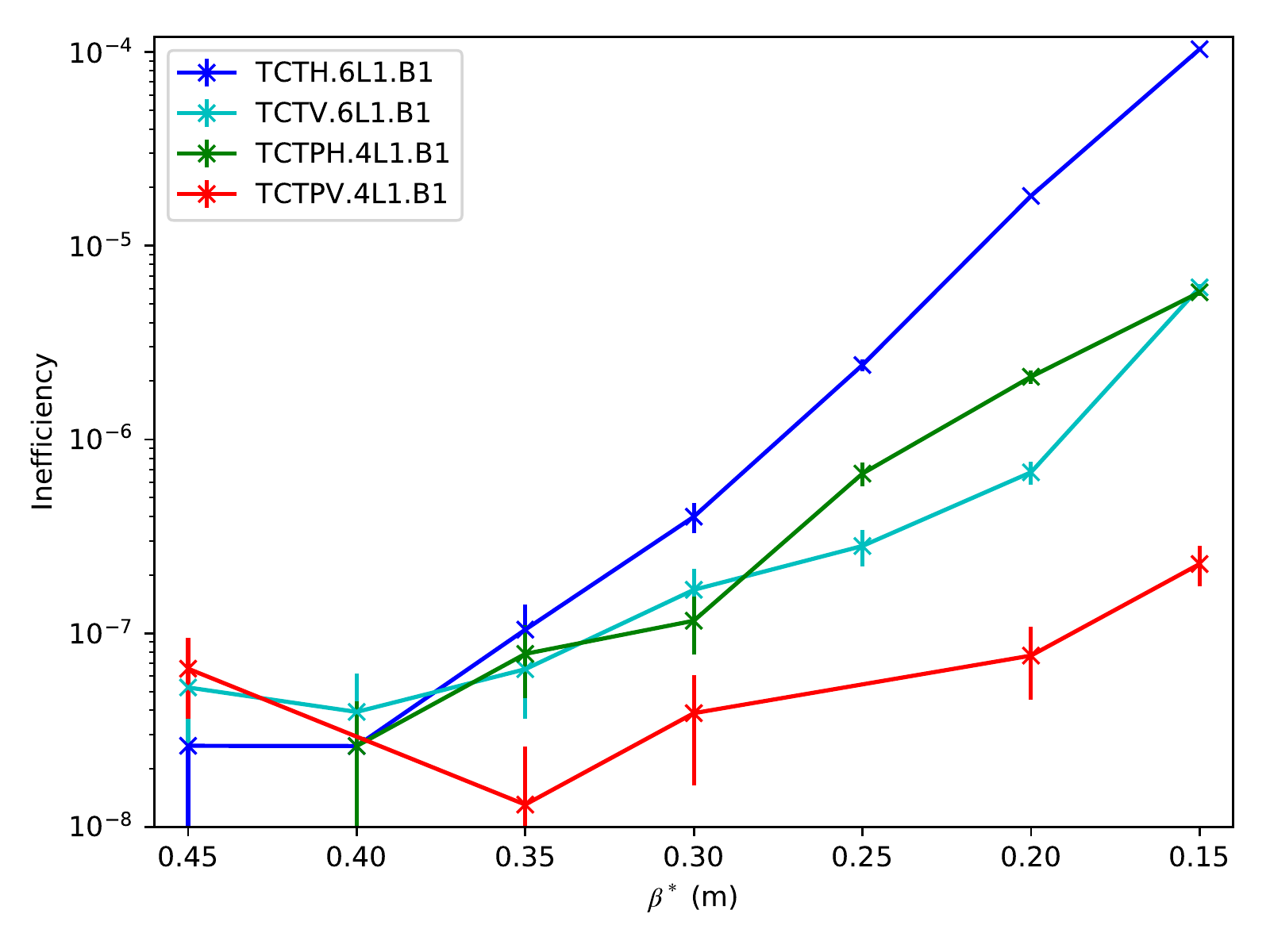}
		\caption{Merlin++ simulated IR1 TCT losses for horizontal excitation as a function of \betastar value during HL-LHC luminosity levelling.}
		\label{fig:merlin_hl_b1_h_ir1}
	\end{center}
\end{figure}

\begin{figure}[!htbp]
	\begin{center}
		\includegraphics[width=1.0\columnwidth]{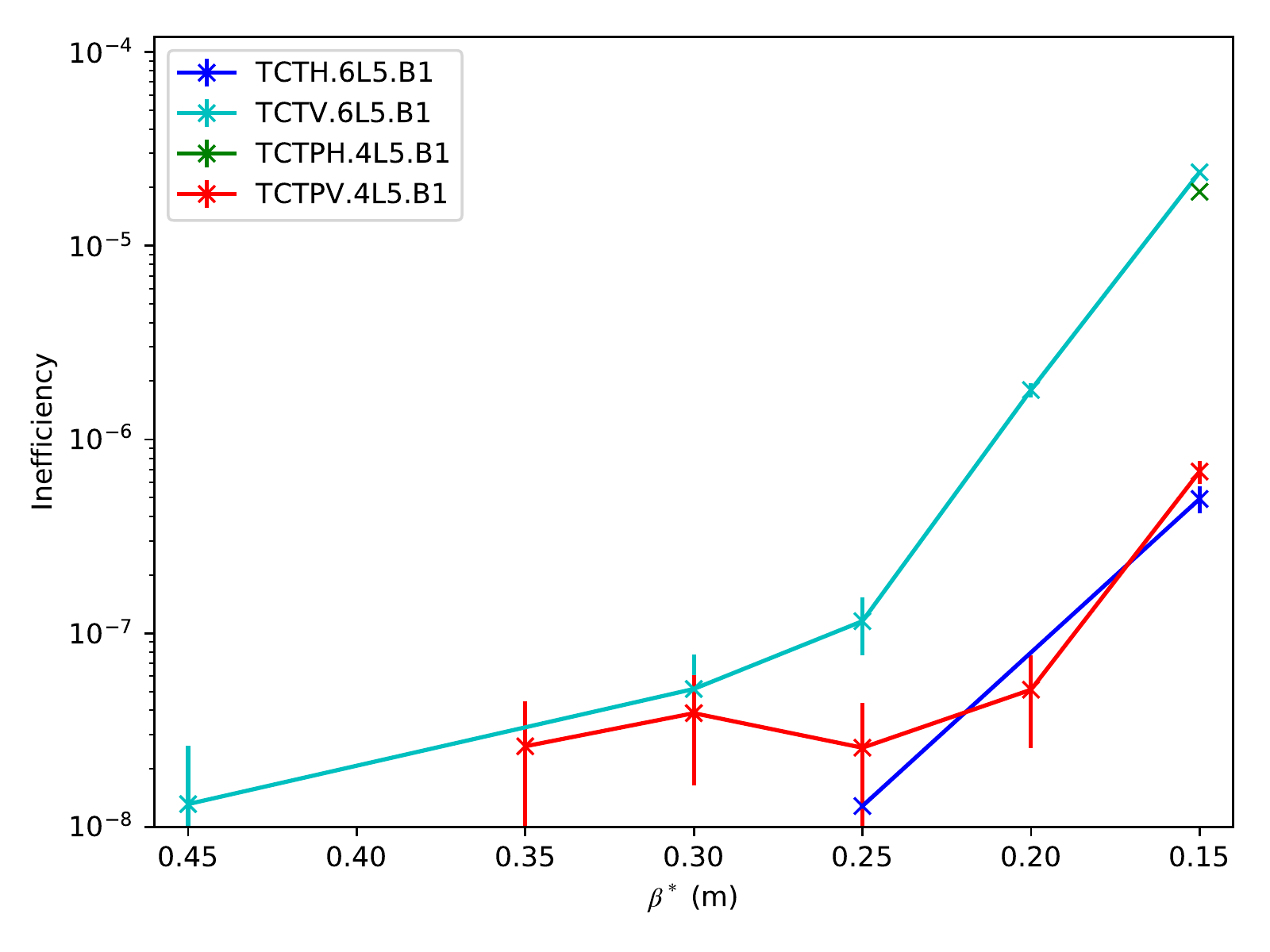}
		\caption{Merlin++ simulated IR5 TCT losses for horizontal excitation as a function of \betastar value during HL-LHC luminosity levelling.}
		\label{fig:merlin_hl_b1_h_ir5}
	\end{center}
\end{figure}

Figures~\ref{fig:merlin_hl_b1_h_ir1} and \ref{fig:merlin_hl_b1_h_ir5} show the Beam 1 losses on the TCTs at the main IPs as a function of the \betastar value.


Losses in cold magnets in the rest of the ring are significantly lower than in the LHC configurations due to the TCLDs catching the dispersive losses. The total cleaning inefficiency of the IR7 collimators at \betastar of 45~cm is found to be \num{7.0e-5} and \num{6.0e-5} for horizontal and vertical excitation respectively, and \num{2.9e-4} and \num{1.3e-4} at 15~cm. For the horizontal 45~cm case, that is that 99.993~\% of lost protons are absorbed in the IR7 collimators.

Although there are only very few direct proton losses in cold elements, the showers from the collimators could potentially still quench magnets during the 1~MW loss scenario, but this has been studied with energy deposition studies in~\cite{lechner_power_2014}.
Studies on the response of the collimators themselves to these loads have been performed in~\cite{bertarelli_dynamic_2018}.

\section{Conclusion}

The LHC collimation system is essential to protect the machine from beam losses during operation. Its performance is continuously monitored by the BLM system. We can use existing measurements to validate simulations, which can then be used to make prediction of future performance for HL-LHC.

In this paper we show that Merlin++ is able to model the proton losses around the LHC. It can reproduce the patterns of losses around the LHC ring and in the interaction regions from measured data. It gives good agreement, to measurements taken with the BLM systems during the beam squeeze, for Run 1 and 2 operation at 4 and 6.5~TeV, both in the overall loss patterns and the changes on the TCTs as the optics configuration is changed. The remaining deviations between simulation and data can be understood by considering the crosstalk between elements due to radiation showers which is not included in Merlin++.

In addition to the SixTrack code, already used successfully for collimation studies, we can therefore also use Merlin++ to predict losses in the future HL-LHC configuration. The possibility to use different simulation tools provides increased flexibility and allows estimating systematic uncertainty in the final results.
We find that the HL-LHC collimation system performs well with a low cleaning inefficiency. The losses on the cold magnets are acceptable, although the loads in the 1 MW scenario imply also the need of energy deposition studies of the magnet coils, as well as thermo-mechanical studies of the most loaded collimators.

\section{Acknowledgements}

We would like to thank Alessio Mereghetti, Rogelio Tomas Garcia and Luis Medina for useful discussion and guidance and Belen Salvachua for BLM data collection.

This work is supported by STFC (UK) grant High Luminosity LHC : UK (HL-LHC-UK), grant number ST/N001621/1.

\FloatBarrier

\bibliography{MyLibrary2,extra}

\end{document}